\documentclass[12pt]{article}

\usepackage{graphics, color}
\usepackage{graphicx}
\usepackage{amssymb}
\usepackage{amsmath}
\usepackage{cite}
\pdfoutput=1

\def\gsim{\, \rlap{$>$}{\lower 1.1ex\hbox{$\sim$}}\,}
\def\lsim{\, \rlap{$<$}{\lower 1.1ex\hbox{$\sim$}}\,}

\def\R{{{\cal R}}}

\def\Z{{{\mathbb Z}}}

\def\Op{{\mathcal{O}}}

\def\tr{{\rm tr}}
\def\ap{\alpha'}
\def\gYM{g_{YM}}

\def\cgs{\cos^2\gamma}
\def\sgs{\sin^2\gamma}

 \newcommand{\be}{\begin{equation}}
\newcommand{\ee}{\end{equation}}
 \newcommand{\bal}{\begin{align}}
 \newcommand{\eal}{\end{align}}
\newcommand{\ben}{\begin{equation*}}
\newcommand{\een}{\end{equation*}}
\newcommand{\bea}{\begin{eqnarray}}
\newcommand{\eea}{\end{eqnarray}}
\newcommand{\bean}{\begin{eqnarray*}}
\newcommand{\eean}{\end{eqnarray*}}
\newcommand{\bes}{\begin{subequations}}
\newcommand{\ees}{\end{subequations}}
\def\p{\partial}
\def\Tr{{\rm Tr}}

\def\eps{{\epsilon}}
\def\tdr{{\tilde r}}
\def\CE{{{\cal E}}}
\def\RR{{\mathbb R}}

\usepackage{epsfig}

\begin{document}

% Title page:

\begin{titlepage}

\begin{flushright}
BRX-TH-637
\end{flushright}

\bigskip
\bigskip\bigskip\bigskip
%\centerline{\Large A gravity dual of axion monodromy}
\centerline{\Large Axion monodromy in a model of holographic gluodynamics}
%\centerline{\Large Axion monodromy in strongly coupled gauge theory}
\bigskip\bigskip\bigskip
\bigskip\bigskip\bigskip

\centerline{{\bf Sergei Dubovsky${}^{1}$, Albion Lawrence${}^{1,2}$, and Matthew M. Roberts${}^{1}$}}
\medskip
\centerline{\em ${}^1$ Center for Cosmology and Particle Physics}
\centerline{\em Department of Physics, New York University}
\centerline{\em 4 Washington Place, New York, NY 10003}
\medskip
\smallskip
\centerline{\em ${}^2$ Martin Fisher School of Physics, Brandeis University}
\centerline{\em MS 057, 415 South Street, Waltham, MA 02454}
\bigskip
\bigskip\bigskip

%ABSTRACT

\begin{abstract}
The low energy field theory for N type IIA D4-branes at strong 't Hooft coupling, wrapped on a circle with antiperiodic boundary conditions for fermions, is known to have a vacuum energy which depends on the $\theta$ angle for the gauge fields, and which is a multivalued function of this angle.  This gives a field-theoretic realization of ``axion monodromy'' for a nondynamical axion.  We construct the supergravity solution dual to the field theory in the metastable state which is the adiabatic continuation of the vacuum to large values of $\theta$. We compute the energy of this state and show that it initially rises quadratically and then flattens out.  We show that the glueball mass decreases with $\theta$, becoming much lower than the 5d KK scale governing the UV completion of this model. We construct two different classes of domain walls interpolating between adjacent vacua.  We identify a number of instability modes -- nucleation of domain walls, bulk Casimir forces, and condensation of tachyonic winding modes in the bulk -- which indicate that the metastable branch eventually becomes unstable.  Finally, we discuss two phenomena which can arise when the axion is dynamical; axion-driven inflation, and axion strings.

\end{abstract}
\end{titlepage}
\baselineskip = 17pt
\tableofcontents
\setcounter{footnote}{0}

\section{Introduction}

Axions are scalars taking values in $S^1$; their periodicity prevents perturbative corrections to the scalar potential, so that their mass is generated entirely by nonperturbative effects and can thus be kept small.  Typically, the axion couples to some nonabelian gauge sector via
\be
	L_{\phi-G} = \frac{\phi}{64\pi^2 f_{\phi}} \tr G  \wedge G
\ee
where $G$ is the field strength for some nonabelian gauge theory, and $\phi \equiv \phi + 2\pi f_{\phi}$.  When the dilute instanton gas approximation is valid, the axion potential is to good approximation:
\be
	V(\phi) = \Lambda^4 \cos \left(\frac{\phi}{f_{\phi}}\right)
\ee
Here $\Lambda$ is the dynamical scale of the gauge theory.  

\begin{figure}[ht!]
\begin{center}
\includegraphics[scale=.85]{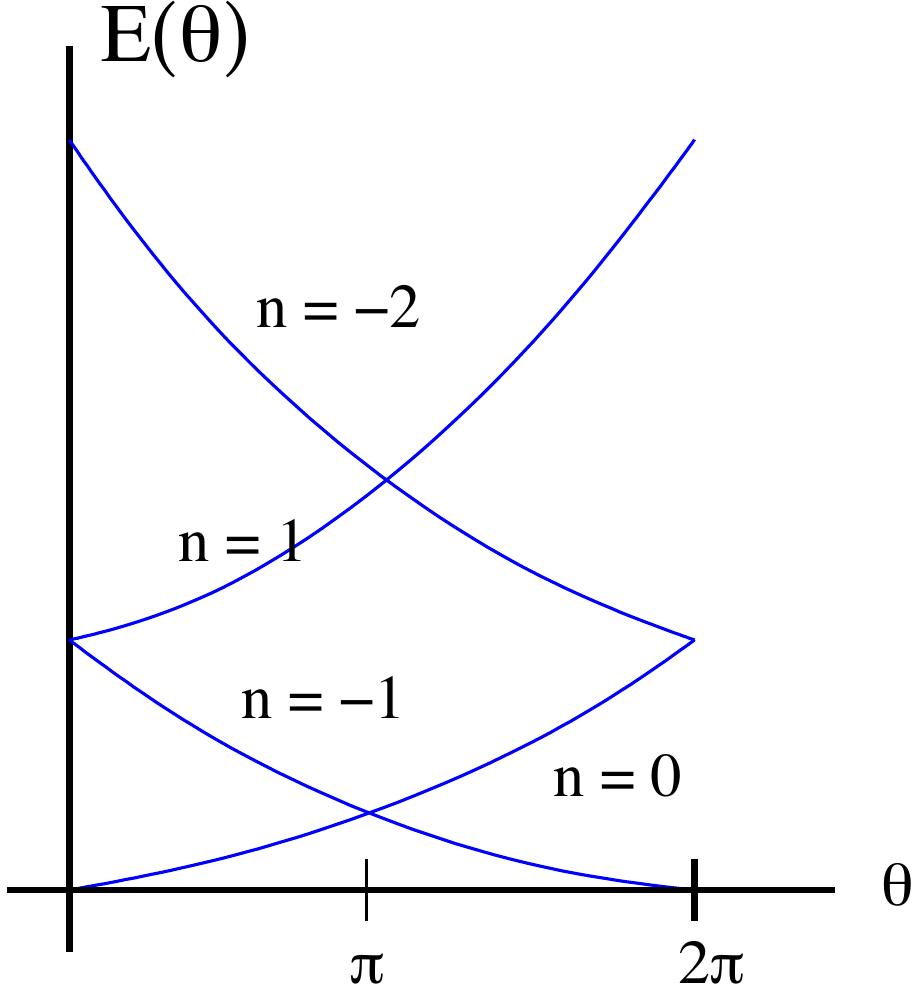}\end{center}
\vspace{-.5cm}
\caption{\label{monofig}The monodromy potential. The spectrum is invariant under shifts of $\theta$ by $2\pi$; a given state, under adiabatic evolution of $\theta$ rises in energy, becoming metastable when $\theta$ shifts by $\pi$ away from its value at zero energy.}
\end{figure}

In general the story is more complicated for confining gauge theories.  This can be seen by studying the dependence of the vacuum energy $V(\theta)$ on the theta term; the angle $\theta$ can be taken to be a nondynamical axion $\theta = \phi/f_{\phi}$ with periodicity $2\pi$.   The dilute gas expansion is known to break down, as seen explicitly in lattice models \cite{Giusti:2007tu}.  In large-N theories with or without fundamental matter \cite{Witten:1980sp,Witten:1998uka}, the energy is a multivalued functional of $\phi$, with a tower of metastable states above the ground state.  While the spectrum is periodic under $\theta \to \theta + 2\pi$, the different states mix and reshuffle, as seen in Figure \ref{monofig}; a given state, under adiabatic change in $\phi$, will continue to increase in energy, initially as $V \sim \theta^2$.  Following a similar story in string theory, \cite{McAllister:2008hb}, we will refer to this phenomenon as ``axion monodromy''.  (We will further abuse the terminology with the phrase ``large values of the axion'', used when we have moved far along a given branch of the axion potential.)

Axion monodromy can lead to interesting phenomena in cosmology and astrophysics.  One application is to building inflation models in field theory and string theory \cite{Silverstein:2008sg,McAllister:2008hb,Kaloper:2008fb,Berg:2009tg,Kaloper:2011jz}.  Models which produce observable gravitational waves, such as simple chaotic inflation models, must have an inflaton which varies over many times the 4d Planck scale in field space \cite{Lyth:1996im,Efstathiou:2005tq}.  Planck-suppressed irrelevant operators in such models tend to spoil slow roll. In axion monodromy inflation, while $f_{\phi} < m_{pl}$, the axion executes many circuits and thus travels over super-Planckian ranges, allowing for models with observable gravitational waves.  In four-dimensional quantum field theory models, the periodicity of the axion keeps quantum corrections under control \cite{Kaloper:2011jz}; string theory models bear out this intuition \cite{Silverstein:2008sg,McAllister:2008hb}.\footnote{These field theory and string theory models are not completely safe; moduli with masses of order the inflatonary Hubble scale or smaller can spoil slow roll \cite{Silverstein:2008sg,McAllister:2008hb,Kaloper:2011jz}.} 

Two features of these models are especially worth noting.  First, in the string theory constructions of \cite{Silverstein:2008sg,McAllister:2008hb,Dong:2010in}, the potential always flattens out at large values of the inflaton (by ``flattens out'' it is generally meant that $V''<0$, equivalently that it asymptotes to a powerlaw $V\sim\phi^\alpha,~\alpha<1$).  This flattening is useful for building inflation models even if inflation does not occur at a high scale. Secondly, these models typically have nonperturbative instabilities in which the inflaton can hop between branches of the potential, or the parameters of the potential can change \cite{Kaloper:2011jz}.  
%In that work it was found that these instabilities, though exponentially suppressed, become less suppressed at large values of the axion expectation value.

Another class of interesting phenomena where axion monodromy may be important is related to a possibility of existence of a plenitude of ultra-light axions, dubbed the axiverse \cite{Arvanitaki:2009fg}.  Even if these axions are coupled to the Standard Model fields purely gravitationally they may lead to observational signatures. First, ultra-light axions would constitute a fraction of dark matter and give rise to a feature in the power spectrum at the scales corresponding to the horizon size at the beginning of axion oscillations.
The size of the effect is proportional to the fraction of the axion component in the dark matter, $\Omega_{ax}/\Omega_{DM}$.

Second, ultra-light axions affect dynamics of astrophysical black holes \cite{Arvanitaki:2009fg,Arvanitaki:2010sy,Dubovsky:2010je}.  Rotating black holes have a superradiant instability in the presence of axions whose Compton wavelength is of order the size of the black hole.  This can lead to gaps in the spectrum of black holes as a function of mass and spin and to a gravity wave signal from the transitions in the axion cloud carrying the initial spin of a black hole. The gravity wave signal is proportional to the fraction of the black hole mass released in the axion cloud, $M_{ax}/M_{BH}$.

In both cases one finds that
\[
{\Omega_{ax}\over \Omega_{DM}},\; {M_{ax}\over M_{BH}}\propto {\Delta\phi^2\over M_{Pl}^2}\ ,
\]
where the effective axion field range $\Delta\phi$ gets enhanced in the presence of monodromy,
\[
\Delta\phi\sim 2\pi N_{mon}f_\phi\;,
\]
and $2\pi N_{mon}$ is the range that $\theta$ has traveled along a given branch of $E(\theta)$. In both cases the monodromy may lead to the significant enhancement of the signal. In particular, if $\Omega_{ax}/\Omega_{DM}$ is of order one as a consequence of the monodromy, the ultra-light axion may play the role of quintessence \cite{Panda:2010uq}.

Yet another implication of monodromy for superradiance is related  to the fact that the production of a large number of axion particles will essentially drive the axion to large values in the vicinity of the black hole horizon.  In QCD-like gauge sectors with light fundamental fermions, a given branch of the axion potential becomes unstable to decay via hidden sector meson production if mass ratios of some flavors are not too large. If this hidden sector couples to the Standard Model,  spectacular electromagnetic signatures may result.  In the case of pure glue, Shifman \cite{Shifman:1998if}\ has speculated that for large enough axion expectation value, the adiabatic continuation of the vacuum will also go from being metastable to being unstable. 

Motivated by these applications, we will study in depth the specific gauge theory constructed in \cite{Witten:1998uka}\ as we increase the theta angle (to be thought of as a nondynamical axion) adiabatically over many periods.  This gauge theory can be constructed as the large-N, large-'t Hooft coupling limit of the theory of massless open strings living on N D4-branes wrapped on a circle with anti-periodic boundary conditions for the fermions breaking supersymmetry. The theory has a supergravity dual, as identified in \cite{Witten:1998uka}\ which makes it amenable to study.  The axion expectation value/theta angle is dual to Ramond-Ramond two-form flux on a specific 2-cycle. Ref. \cite{Witten:1998uka}\ studied this solution for small values of the two-form flux, in which backreaction on the geometry can be ignored. We will find the full solution including backreaction for large values of $\theta$.\footnote{This was attempted once before in \cite{Barbon:1999zp}; as we will discuss in \S2, that background does not satisfy the supergravity equations of motion.}  In \S3\ and \S4 we then go on to study the physics of this model at large $\theta$. In \S5\ we discuss some phenomena that can arise when the axion becomes dynamical. In \S6\ we conclude with some open questions.

\subsection{Brief summary of results}

The system we study is surprisingly rich, and reflects the phenomena described above in other realizations of axion monodromy.  Before launching into a detailed examination, let us highlight the qualitative results.

First, the energy as a function of $\theta$ flattens out considerably at large $\theta$; similarly, the energy gap between branches becomes smaller.  The theory has a deconfined phase whose energy sets an upper bound on the energy of confining vacua as a function of $\theta$.   

A reader aware of the history of this model will recall that it is poor for studying QCD for small values of $\theta$, as the Kaluza-Klein scale at which the theory becomes $4+1$-dimensional is close to the deconfinement scale.  However, we will find that for large $\theta$ the mass gap of the theory will occur at lower and lower energies, well below the KK scale. 

We will construct solutions for axion domain walls interpolating between different values of $\theta$ (at small values of $\theta$ these were identified in \cite{Witten:1998uka}.)  These arise as various configurations of D6-branes wrapping internal cycles of the 10d dual geometry.  We compute the scaling of their tensions as a function of $\theta$.  We will show that there is a nonperturbative instability that has a supergravity description closely analogous to \cite{Kachru:2002gs}.  As $\theta$ gets large, the potential barrier between branches gets very small, until the metastable vacuum becomes unstable.  We will identify additional modes of instability directly in the supergravity solution, namely the appearance of tachyonic string modes and the dominance of Casimir forces which appear to render the vacua at large $\theta$ unstable.  We will compare the values of $\theta$ as a function of $N,\lambda$ at which these effects appear. Finally, we will consider the physics that arises when $\theta$ is promoted to a dynamical axion.  We will identify axion strings which can bound domain walls, and we will show that this field theory provides an interesting model of inflation at a comparatively low scale ({\it eg} with unobservable tensor modes).

\section{The dual pair}

In this work we revisit the field theory first described in \cite{Witten:1998zw,Witten:1998uka}, which arises as the low-energy dynamics of N type IIA D4-branes wrapped on an $S^1$ with periodicity $2\pi\beta$ and antiperiodic boundary conditions for spacetime fermions.  At tree level, this leaves us with an $SU(N)$ theory with adjoint scalars, which is four-dimensional in the IR and which has a five-dimensional UV completion at energies of order $\beta$.  It is expected that the scalars will get masses at the loop level, which will be large at large 't Hooft coupling, so that the theory is pure Yang-Mills at energies below the 5d KK scale.  

The tree level couplings of $N$ D4-branes can be computed from the combined Dirac-Born-Infeld plus Chern-Simons actions\footnote{We will use the normalization conventions of \cite{Polchinski1998string} in this paper.}, which we list here for general $Dp$,
\be
S_{Dp}=S_{DBI}+S_{CS},
\ee
where
\be
S_{DBI}=-\mu_p\int d^{p+1}\xi \Tr \left\{e^{-\phi} \left[ -\det\left( g_{\mu\nu}+B_{\mu\nu}+2 \pi \alpha' F_{\mu\nu}\right)\right]^{1/2}\right\},\label{eq:dbiaction},
\ee
and
\be
S_{CS}=\mu_p\int_{p+1}\Tr\left[ e^{2\pi\alpha' F+B}\wedge\sum_q C_q\right],~\mu_p=\frac{1}{(2\pi)^p\ap^{(p+1)/2}}\ .
\label{eq:csaction}
\ee
Here $F$ is the world-volume Yang-Mills field, and the traces are taken in the fundamental representation. In practice this is a bit of a cheat since the nonabelian DBI action is not known.  However, we are really only interested in the quadratic term, which will survive the decoupling limit.
The D-brane tension is 
\be
\mu_p=\frac{1}{(2\pi)^p\ap^{(p+1)/2}}\ ,
\ee
where $g_s=\langle e^\phi \rangle$; and the string tension is $T = 1/(2\pi \ap)$. Expanding (\ref{eq:dbiaction}) to quadratic order in $F$, we find that the 5d gauge kinetic term is 
\be
	S_{kin} = \frac{1}{16\pi^2 g_s \sqrt{\alpha'}}\int d^5 x \Tr F^2
\ee
Thus the 5d gauge coupling is $g_5^2 = 4\pi^2 g_s \sqrt{\alpha'}$. Now wrap the D4-branes on a compact circle parameterized by $\chi$ with proper length $2 \pi\beta$, and consider a constant Wilson line $C = C_\chi d\chi$ for the Ramond-Ramond (RR) one-form. Expanding (\ref{eq:dbiaction},~\ref{eq:csaction}) to quadratic order, we then have
\be
S =\frac{\beta}{8\pi\sqrt{\ap}g_s}\int_{\R^3+1}\Tr F^2+\frac{C_\chi\beta}{4\pi\sqrt{\ap}}\int_{\R^{3+1}} \Tr F\wedge F\ . \label{eq:dimredact}
\ee
The standard gauge theory parameters are defined via the quadratic action:
\be
=\frac{1}{4\gYM^2}\int_{3+1}\tr F^2+\frac{\theta}{8\pi^2}\int_{3+1}\tr F\wedge F\ .\label{eq:gtnorm}
\ee
For Euclidean instantons, $\int F\wedge F = 8\pi^2 n$, $n \in Z$, so that with this normalization,  $\theta \equiv \theta + 2\pi$, and we can identify:
\be
g_{YM}^2 = \frac{g_5^2}{2\pi \beta}=\frac{2\pi g_s\sqrt{\ap}}{\beta},~\theta = \frac{2\pi C_\chi\beta}{\sqrt{\ap}}.\label{4dparam}
\ee

We will also be interested in the M-theory lift of this solution.  The Ramond-Ramond one-form in type IIA descends from the 11d metric $G_{z\mu}$ where $z$ denotes the circle on which one reduces M theory to type IIA string theory, and $\mu = 0,\ldots 9$.  In 11 dimensions, this means that the $z-\chi$ torus has a complex structure with a nontrivial real part.  As we will see when studying the backreacted M5s, we will be wrapping the branes on a torus with canonical form
\be
K^2|d\sigma_1+\tau d\sigma_2|^2,~\tau=\frac{\theta}{2\pi}+\frac{2\pi i}{g_{YM}^2}
\ee
For $M5$-branes wrapping this torus, the complex structure $\tau_{T^2}$ is equal to the complexified gauge coupling $\tau_{YM}$ defined from $\gYM$ and $\theta$ in (\ref{4dparam}). For SUSY boundary conditions in both $\sigma_a$ we would have a full $SL(2,\mathbb{Z})$ symmetry.  Absent supersymmetry, only the T transformation $\tau \rightarrow \tau+1$, which corresponds to shifts $\theta \rightarrow \theta + 2 \pi$, is guaranteed to be a symmetry of the theory.

We now wish to study the near-horizon limit of these solutions in the limit that $N$ grows with fixed $g_5^2,\beta,\theta$.  Because of the anti-periodic boundary conditions, there are two possibilities for the near-horizon geometry \cite{Witten:1998zw,Witten:1998uka}.  One possibility is that the tree-level solution is identical to that of the supersymmetric D4-brane solution.  In this case, the circle $S^1$ degenerates only at the Poincar\'e horizon.  The RR 1-form can be constant everywhere and does not backreact on the geometry; the physics is independent of $\theta$.  We will call this the ``deconfined'' phase as the solution is well defined down to the Poincar\'e horizon, so that there is no mass gap and no area law behavior for the Wilson lines. There is, however, a conical singularity at the Poincar\'e horizon, and the solution is known to be unstable to winding string tachyons \cite{Horowitz:2005vp}.

In the lower-energy solution \cite{Witten:1998zw,Witten:1998uka}\ the $S^1$ pinches off at a finite distance in the radial direction away from the D4-branes, indicating confinement and a mass gap.  The $S^1$ together with the radial direction have the topology of a disc, and a finite Wilson line for $C$ about the $S^1$ then requires a nonvanishing RR 2-form flux through the disc.  In this case, the vacuum energy is $\theta$-dependent.  We will call this the ``confined'' phase. We wish to find the corresponding solution for arbitrary RR 2-form flux, at least in the regime where the solution remains valid.  We will begin by studying the solutions in M-theory. 

\subsection{M theory solution}

The 11d Euclidean solution for N non-extremal M5-branes is \cite{Gueven:1992hh,Duff:1999rk}:
\begin{eqnarray}
	ds^2 & = & \frac{\Delta_+}{\Delta_-^{2/3}} d\chi^2 + \Delta_- \left(dt_E^2 + d\vec{x}_3^2 + dz^2\right) + \frac{1}{\Delta_+ \Delta_-} d\tdr^2 + \tdr^2 d\Omega_4^2\nonumber\\
	F_4 & = & q_m \eps_{S^4}\label{eq:blackmfive}
\end{eqnarray}
where
\be
	\Delta_{\pm} = 1 - \frac{r_{\pm}^3}{\tdr^3}\ ;
\ee
$d\Omega_4$ is the metric on the 4-sphere with unit radius; and $\eps_{S^4}$ is the corresponding volume form. The parameters $r_{\pm}$ satisfy the constraint
\be
	(r_+r_-)^3 = \left(\frac{q_m}{9}\right)^2 \equiv \left(\frac{\ell_{11}^3 N}{8\pi}\right)^2
\ee
where $N$ is the number of M5-branes, and Newton's constant in 11 dimensions is $2\kappa_{11}^2 = \left(2\pi\right)^8 \ell_{11}^9$.   Conventionally we take $r_+ \geq r_-$.  $F_{4}$ is the four-form field strength.  

In (\ref{eq:blackmfive}), $\chi$ is usually taken to be Euclidean time, with antiperiodic boundary conditions for the fermions, and $r_0^3 = r_+^3 - r_-^3$ is related to the temperature of the black M5-brane.  In this paper, we are interested in the Lorentzian solution at zero temperature for an M5-brane wrapping a circle with the same boundary conditions for the fermions.  In passing to Minkowski signature we will  choose to analytically continue $t_E \to -i t$; this will preserve the fact that (\ref{eq:blackmfive}) is a solution to the equations of motion.

In order to describe a solution to the equations of motion with $G_{z\chi}\neq 0$, we perform a Euclidean rotation by an angle $\gamma$ in the $z-\chi$ direction, 
\be\label{eq:eucrot}
	z \to z \cos \gamma  - \chi \sin\gamma \ ;\ \ \chi \to \chi \cos\gamma + z \sin\gamma\ .
\ee
This is just a change of coordinates, so that the transformation of (\ref{eq:blackmfive}) is guaranteed to produce a solution to the equations of motion with a nonvanishing value of $G_{z\chi}$.  Next, we define $r^3 \equiv \tdr^3 - r_-^3$ and $r_0^3 = r_+^3 - r_-^3$.  The resulting M theory metric is\footnote{We have also rescaled various coordinates so that the IIA reduction is asymptotically flat with the standard normalization.}:
\begin{eqnarray}
ds^2_{11} & = & e^{-2\phi_0/3}\left[\frac{1}{H_4^{1/3}}\left(-dt^2+dy_i^2 +\frac{f}{H_0}d\chi^2\right) +H_4^{2/3}\left( \frac{dr^2}{f}+r^2d\Omega_4^2\right)\right]\nonumber\\
& & \ \ \ \ \ +\frac{e^{4\phi_0/3}H_0}{H_4^{1/3}}\left(dz-\frac{e^{-\phi_0}}{H_0} \left(1 - f\right) \sin\gamma\cos\gamma d\chi \right)^2\nonumber\\
F& = & 3\pi N \ell_{11}^3\epsilon(\Omega_4),\label{flatMsolution}
\end{eqnarray}
where 
\be
H_4=1+\frac{c_4^3}{r^3},~f=1-\frac{r_0^3}{r^3},~H_0= 1-\frac{r_0^3\sgs}{r^3},~c_4^3=\frac{\sqrt{4\pi^2 \ell_{11}^6 e^{2\phi_0}N^2+r_0^6}-r_0^3}{2}\ ;
\ee
$\epsilon(\Omega_4)$ is the volume form on the four-sphere. The awkward normalization of the coordinates is such that the IIA reduction, asymptotes to the canonical flat metric in string frame. At this stage we have not yet compactified the $z$ direction (avoiding a singularity, before or after implementing (\ref{eq:eucrot}), requires compactifying $\chi$ with specific period).  We are therefore free to make a further diffeomorphism of $z$, $z \to z + e^{\phi_0} \tan\gamma$, so that the metric in (\ref{flatMsolution}) becomes
\begin{eqnarray}
ds^2_{11} & = & e^{-2\phi_0/3}\left[\frac{1}{H_4^{1/3}}\left(-dt^2+dy_i^2 +\frac{f}{H_0}d\chi^2\right) +H_4^{2/3}\left( \frac{dr^2}{f}+r^2d\Omega_4^2\right)\right]\nonumber\\
& & \ \ \ \ \ +\frac{e^{4\phi_0/3}H_0}{H_4^{1/3}}\left(dz + \frac{f}{e^{\phi_0}H_0} \sin\gamma\cos\gamma d\chi \right)^2
\end{eqnarray}
This will ensure that the RR one-form in the type IIA reduction is nonsingular\footnote{By nonsingular we mean we wish to work in a gauge where $C_\mu C^\mu<\infty$, which is a gauge condition. All Ramond-Ramond field strengths are smooth.} at $r = r_0$.

We also can now explicitly verify that on compactifying down to 8+1 dimensions, we find $\tau_{T^2}=\tau_{YM}$ Let $z$ have periodicity $2\pi \ell_{11}$, where the 11d Newton's constant is $2\kappa_{11} = (2\pi)^8 \ell_{11}^9$, and proper radius $R_{11} \equiv e^{2\phi_0/3} \ell_{11}$, which defines $\phi_0$; this will become the type IIA dilaton.  If we normalize these coordinates as coordinates $\sigma_1=z/\ell_{11},~\sigma_2=\chi/\beta$, the asymptotic torus metric is
\be
ds^2_{T^2}=e^{4\phi_0/3}\ell_{11}^2d\sigma_1^2+2e^{2\phi_0/3}\ell_{11}\beta \tan\gamma d\sigma_1 d\sigma_2  + e^{-2\phi_0/3}\beta^2 \sec^2\gamma d\sigma_2^2\ ,
\ee
which takes the canonical form $K^2|d\sigma_1+\tau d\sigma_2|^2$, with
\be
\tau_{T^2} = \frac{\beta\tan\gamma}{e^{\phi_0}\ell_{11}}+i\frac{\beta}{e^{\phi_0}\ell_{11}},~K=e^{2\phi_0/3}\ell_{11},~\ap=\ell_{11}^2.
\ee

\subsection{Type IIA solution}

After performing the Euclidean rotation, we compactify both $z$ and $\chi$.  $z$ is compactified with coordinate periodicity $2\pi\ell_{11}$, so that the physical radius is $R_{11} = e^{2\phi/3} \ell_{11}$. The radius $\beta$ of the $\chi$ circle will be fixed by regularity of the solution.  The type IIA string theory metric $ds_{IIA}^2$, the RR 1-form $C_{\mu}$, and the dilaton $\phi$ are found by writing the M theory metric in the form
\be
ds^2_{11}=e^{-2\phi/3}ds_{IIA}^2 +e^{+4\phi/3}(dz+C_\rho dx^\rho)^2,\label{KKred}
\ee
Applying this to (\ref{flatMsolution}), we find the string frame metric:
\be
ds^2_{IIA} = \sqrt{\frac{H_0}{H_4}}\left(-dt^2+dy_i^2 \right)+\frac{1}{\sqrt{H_4H_0}}d\chi^2+\sqrt{H_4H_0}\left(\frac{dr^2}{f}+r^2d\Omega_4^2 \right)\ ,\label{eq:IIAafsol}
\ee
with a nontrivial dilaton:
\be
e^{\phi}=e^{\phi_0}\frac{H_0^{3/4}}{H_4^{1/4}}
\ee
and Ramond-Ramond 2- and 4-form field strengths:
\begin{eqnarray}
	F^{(2)} & = & dC^{(1)}\nonumber\\
	C^{(1)} & = & e^{-\phi_0}\frac{f}{H_0}\sin\gamma\cos\gamma d\chi\nonumber\\
	F^{(4)} & = & 3\pi N{\alpha'}^{3/2} \epsilon_4\ .
\end{eqnarray}
Here $H_{0,4}$ are defined as above, with the factor of $\ell_{11}^6$ in $c_4^3$ replaced with $(\alpha')^3$.  $\phi_0$ is the asymptotic value of the dilaton, and $\Omega_4^2$ and $\epsilon_4$ are the line element and volume form on a unit $S^4$.  

The reader who is used to statements like $L_{11} = g_s^{1/3} \sqrt{\alpha'}$, where $L_{11}$ is the 11d Planck scale, may be puzzled by this simple substitution in $c_4$ and in the definition of the Ramond-Ramond four-form $F$.  The point is that the 11d Planck scale in the Einstein frame of 11d supergravity is the string scale in the string frame of 10d type IIA supergravity. A fuller explanation of this fact is provided in Appendix A.

The $r-\chi$ submanifold has the topology of a disc with the center at $r = r_0$. One  must adjust the parameters to ensure regularity of the metric at $r = r_0$.  For our purposes, $\beta$ is a parameter of the D4-brane field theory which we wish to vary at will; regularity then fixes $r_0$ as a function of $\beta$ and $\tan\gamma$:
\be
\chi\sim\chi+2\pi\beta,~\beta=\frac{2\sqrt{H_4(r_0)H_0(r_0)}}{f'(r_0)}.\label{beta1}
\ee
If we set $\tan\gamma \equiv x$, we can write
\be
\beta = \frac{2 r_0}{3\sqrt{1+x^2}}\sqrt{\frac{1}{2}+\sqrt{\frac{1}{4}+\frac{\lambda^2\beta^2\ap^2}{4r_0^6}}}
\ee
We can now fix $r_0$ and $\tan\gamma$ completely in terms of the 4d theta angle and gauge coupling, by plugging our supergravity solution for $C_{\chi}$ and $e^{\phi}$ at $r\to\infty$ into (\ref{4dparam}),
\be
\theta=\frac{2\pi\beta\tan\gamma}{e^{\phi_0}\sqrt{\ap}},\ee
or
\be \tan\gamma=\frac{\gYM^2\theta}{4\pi^2}=\frac{\lambda\theta}{4\pi^2N}\equiv x\ ,\label{thetarel}
\ee
where $\lambda = \gYM^2N$. We will find that the natural variable in the quantum field theory is not $\theta$ but $x$; all of the $\theta$-dependence of the supergravity solution is captured by this variable.

Our solution differs from that given in \cite{Barbon:1999zp}\ in various ways.  We have checked and found that the solution in that paper does not solve the type IIA supergravity equations of motion.  The authors of \cite{Barbon:1999zp}\ performed a double analytic continuation of the smeared black D0-D4-brane solution.  Their solution leaves a complex RR 1-form.  The difference is non-trivial; for example, when we computed the energy as a function of $\theta$ using the solution in \cite{Barbon:1999zp}, we found that it diverged at a finite value of $\theta$ (which is what alerted us to the possibility that this other solution was incorrect).  

\subsection{Decoupling limit}

The next step is to take the low-energy limit while fixing the parameters of the QFT, following \cite{Maldacena:1997re,Itzhaki:1998dd}.  We begin by setting $r \equiv \alpha' u$, $r_0 \equiv \alpha' u_0$.  Note that $u, u_0$ have units of energy.  We then send $\alpha'\to 0$, $g_s = e^{\phi_0} \to \infty$ while keeping $u, u_0, \beta, \theta,\lambda$, and $N$ fixed.  The resulting geometry in type IIA string theory is:
\begin{eqnarray}
ds^2_{IIA} & = & \ap \left[ \sqrt{\frac{2 u^3H_0}{\lambda\beta}}\left(-dt^2+dy_i^2+\frac{f}{H_0}d\chi^2 \right)+\sqrt{\frac{\lambda\beta H_0}{2u^3}}\frac{du^2}{f}+\sqrt{\frac{\lambda\beta  H_0 u}{2}}d\Omega_4^2
 \right],\nonumber\\
C^{(1)} & = & \frac{\sqrt{\ap}2\pi Nf\tan\gamma }{\lambda \beta H_0}d\chi\nonumber\\
e^\phi & = & \frac{1}{\pi N}\left( \frac{\beta\lambda uH_0}{2}\right)^{3/4}\nonumber\\
	 F^{(4)} & = & 3\pi\ap^{3/2}N\epsilon_4\label{D4throat}
\end{eqnarray}
where  $d\Omega^2_4$ and $\epsilon_4$ are the line element and volume form of a unit $S^4$ and 
\begin{eqnarray}
	\lambda & = & \gYM^2 N\nonumber\\
	 f(u) & = & 1-\frac{u_0^3}{u^3}\nonumber\\
	 H_0(u) & = & 1-\frac{u_0^3}{u^3}\sgs.
\end{eqnarray}
In this limit,  (\ref{beta1}) reduces to
\be
\beta=\frac{2\lambda \cgs}{9u_0}=\frac{2\lambda}{9u_0(1+x^2)}\label{betascaled}
\ee
Again, we emphasize that $\theta$ appears in the solution through the variable $x$.

At fixed $\beta$, $u_0 \sim \frac{1}{1 + x^2}$ and becomes small at large $x$.  As we will see below, this corresponds to a lowering of the mass gap of the theory.  The $S^4$ also gets smaller with $u$, and the minimum size shrinks. Similarly, the dilaton becomes weak in the IR, and the string coupling at the tip $u = u_0$ decreases at large $\theta$.  

\subsection{Trustworthiness of the type IIA solution}

Because various cycles are becoming small at large $\theta$, we must check that the type IIA solution is trustworthy; specifically, that the curvature remains small and the string coupling weak in the interior.  This is especially important since we have broken supersymmetry and have no nonremormalization theorems to protect us.

In string frame, the metric curvature is greatest at the tip $u=u_0$, at which point
\be
R^{(10)}=-\frac{27(1+4x^2+3x^4)}{\lambda\ap},~R^{(10)}_{abcd}R_{(10)}^{abcd}=\frac{54}{\lambda^2\ap^2}(1+x^2)^2(13+27x^2+27x^4).
\ee
If $x > \lambda^{1/4}$, the curvatures become string scale.  The radius of the four sphere is $\sqrt{\alpha'}$ at this bound as well:
\be
V(S^4)=\ap^2\frac{\lambda^2}{9(1+x^2)^2}\Omega_4
\ee
It is possible (though we do not have proof) that the solutions will cease to exist at this point, indicating the lack of a stable confined-phase solution for $x > \lambda^{1/4}$.  At any rate, we will find below that stringy winding modes will have this effect when $x > \lambda^{1/3} > \lambda^{1/4}$.

The local string coupling, $e^\phi$ grows as $(H_0(u) u)^{3/4}$ from a minimum at the tip $u = u_0$, where it takes the value:
\be
e^{\phi(u_0)}=\frac{1}{\pi N}\left( \frac{\lambda}{3(1+x^2)}\right)^{3/2}
\ee
This is weak when $\lambda^{3/2} \ll N$, or $g^{3}_{YM} << N^{-1/2}$, which may or may not be true depending on the details of our large-$N$, large-$\lambda$ limit (for example, the coupling will remain small if we take $N$ to be large at fixed $\lambda$.) Even when this limit does not hold, the coupling will be weak when  $x >> \lambda^{1/2}/N^{1/3}$.  The type IIA solution will cross over to M theory in the UV, when $e^\phi\sim 1$. If we assume that $H_0\sim 1$ in this regime, then the crossover occurs at
\be
u_{cross}\approx\frac{2}{\beta \lambda}(\pi N)^{4/3}.
\ee
We will now turn to the decoupling solution in M theory.

\subsection{M theory lift of decoupled geometry}

The M theory lift of our solution in the decoupling limit can be found either by directly taking the scaling limit of (\ref{flatMsolution}), or by inserting the solution (\ref{D4throat}) into (\ref{KKred}).  The resulting metric is:
\begin{eqnarray}
ds^2_{11} & = & \ap(\pi N)^{2/3}\left[\frac{2u}{\beta\lambda} \left(-dt^2+dy_i^2+\frac{f}{H_0}d\chi^2 \right) +\frac{du^2}{f u^2}+d\Omega_4^2\right]\nonumber\\
 & & \ \ \ \ +\frac{\beta\lambda u H_0}{2(\pi N)^{4/3}}\left(dz+\frac{2\pi N\sqrt{\ap}f}{\beta \lambda H_0}\tan\gamma d\chi\right)^2\nonumber\\
 F^{(4)}&=&3\pi\ap^{3/2}N\epsilon_4\ .
 \label{eq:Mdecouple}
\end{eqnarray}
This is the  $AdS_7$ soliton with two spacial field theory directions compactified on a tilted torus. The unusual $N$ scaling in the metric comes from our choosing to work with $\lambda$ instead of $\gYM$. 

The Ricci scalar $R^{(11)}=\frac{3}{2\ap (\pi N)^{2/3}}$ is everywhere constant, and well below string scale in the large $N$ limit. For M theory to be a valid description we require $e^\phi \gg 1$ which corresponds to $\lambda/N=\gYM^2\gg1$.

%\subsection{Undecoupling: Making the axion dynamical}

\section{Field theory spectrum and dynamics}

\subsection{The potential $E(\theta)$.}

Beginning with (\ref{eq:IIAafsol}), we can compute the energy of the solution using the standard ADM expression for asymptotically flat spacetimes (we should be careful that we use as the reference solution flat space with the same value of $\beta$, $e^{\phi_0}$ as our solution):
\be
\frac{E}{V_3}=\frac{3\sqrt{4\pi^2e^{2\phi_0}(\alpha')^3 N^2+r_0^6}-r_0^3}{4\kappa_{10}^2e^{2\phi_0}}2\pi\beta\Omega_4,\label{flatenergy}
\ee
where $\Omega_4=8\pi^2/3$ is the volume of a unit four-sphere, and $2 \kappa_{10}^2 = (2\pi)^7 (\alpha')^4$. When $r_0 =0$, (\ref{eq:IIAafsol}) is just the deconfined solution with an energy independent of $\theta$,
\be
\frac{E(r_0=0)}{2\pi\beta V_3}=\frac{\mu_4}{g_s},\label{D4massregular}.
\ee

The energy $V(\theta)$ in the field theory can be found by taking the scaling limit of (\ref{flatenergy}), or by computing it directly from the metric (\ref{eq:Mdecouple}) following \cite{Horowitz:1998ha}. In the latter case, we have computed the energy of the M theory throat (\ref{eq:Mdecouple}) using the formalism of \cite{Hawking:1995fd,Horowitz:1998ha}.  The formula is:
\be\label{eq:hmenergy}
\frac{E(\theta)}{V_5}=-\frac{1}{\kappa_{11}^2}\oint N(K-K_0)\ ,
\ee
where the integral is over a spacelike slice of the boundary at fixed $t$, $K$ is the trace of the extrinsic curvature of this boundary surface, and $K_0$ is the extrinsic curvature of a spacelike boundary surface with the same intrinsic geometry, in a reference background, which in our case we take to be simply exact planar $AdS_7\times S^4$.\footnote{Specifically, as discussed in \cite{Horowitz:1998ha}, one considers a boundary at {\it finite} radius $u_{UV}$, matches the proper periodicities of $\chi$ and $z$ to the proper periodicities at finite radius in the reference background, and computes $K-K_0$ as a function of this "cutoff" radius.  Only then does one take the radius to infinity.  
%Furthermore, the reference background has a singularity at the horizon due to the compactification of $\chi$ and $z$; as in \cite{Horowitz:1998ha}, we ignore this subtlety and assume the singularity does not contribute to the energy.
}  

The result from the decoupling limit of (\ref{flatenergy}) is:
\be
\frac{E}{V_3}=\frac{N}{4\pi^2\gYM^2\ap^2}-\frac{N^2u_0^3}{12\pi^2\lambda^2\beta}+\Op(\ap).
\ee
The first divergent term is the just the energy of $N$ D4s, which we subtract off.  When we calculate the energy using (\ref{eq:hmenergy}), we find the same expression without this first term (as it cancels out when we subtract the extrinsic curvature of the reference geometry).  The finite term written in terms of field theory quantities via (\ref{thetarel},~\ref{betascaled}) is
\be
\frac{E}{V_3}=-\frac{2\lambda N^2}{3^7\pi^2\beta^4}\frac{1}{\left(1+(\lambda \theta/4\pi^2 N)^2 \right)^3}.\label{thetapotential}
\ee
This formula is valid in both the M theory and IIA regimes, as it is a classical conserved charge.

Eq. (\ref{thetapotential}) has the form $E(\theta) = \lambda N^2 v(\lambda \theta/N)$ argued for in \cite{Witten:1980sp,Witten:1998uka}\footnote{These papers do not specifically address the $\lambda$ dependence of the energy.}.
For small $x=\lambda \theta/4\pi^2 N$ we find the quadratic behavior found in \cite{Witten:1998uka}:
\be
\frac{E}{V_3}=-\frac{2\lambda N^2}{3^7\pi^2\beta^4}+\frac{\lambda^2\theta^2}{2^33^6\pi^6\beta^4}+\ldots\label{quadratictheta}
\ee
As we adiabatically increase $x$ to stay on a fixed branch of $E(\theta)$, however, the potential flattens out considerably, much as in the string theory examples in \cite{Silverstein:2008sg,McAllister:2008hb,Dong:2010in}.  The point is that as we increase $\theta$, the $u-\chi$ throat grows longer and longer (as $u_0$ gets smaller). In the bulk, the solution approaches that of the deeconfined phase, for which the physics is $\theta$-independent.  In the field theory, the mass gap is decreasing with $u_0$, as we will confirm below.  The mass gap is the natural scale governing the change in energy with $\theta$.

\begin{figure}[ht!]
\begin{center}
\includegraphics[scale=.85]{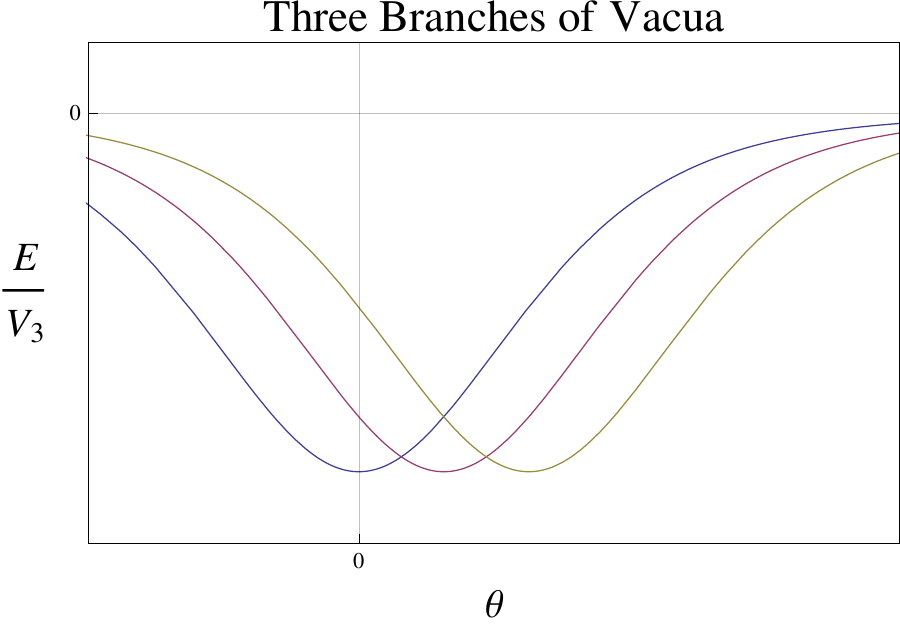}\end{center}
\vspace{-.5cm}
\caption{\label{three_branches}A plot of $E(\theta)$.  The three branches corresponds to shifting $\theta$ by $2\pi$.}
\end{figure}

\begin{figure}[ht!]
\begin{center}
\includegraphics[scale=.85]{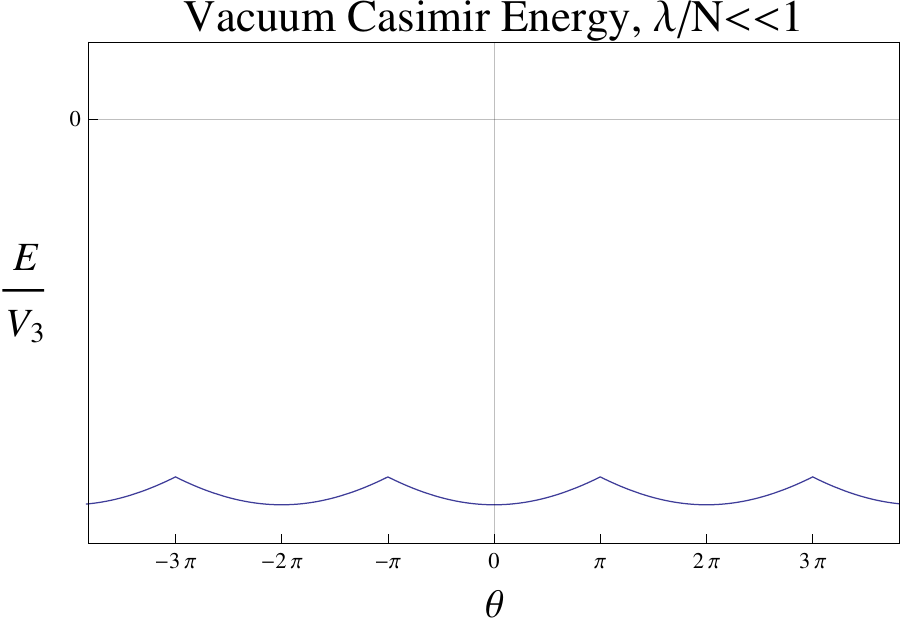}
\includegraphics[scale=.85]{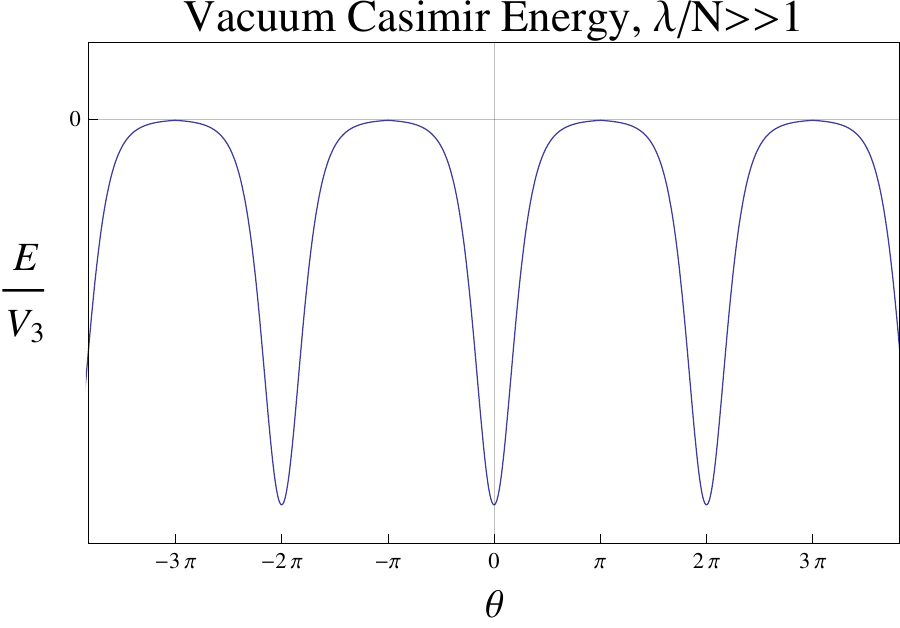}\end{center}
\vspace{-.5cm}
\caption{\label{vacstr} The vacuum energy of the theory. On the left we have the vacuum structure when $\lambda/N \ll 1$ and so the energy behaves quadratically. This is the regime where the IIA description is valid. On the right we have the structure when $\lambda/N \gg 1$, and there is strong departure from the quadratic behavior. This corresponds to the M theory description being appropriate.}
\end{figure}

There are an infinite sequence of branches with the same energetics as (\ref{thetapotential}), save for the substitution $\theta \rightarrow \theta+2\pi n$; this can be seen in the formula for the vacuum energy (following \cite{Witten:1998uka}):
\be
\frac{E}{V_3}=\min_{k\in \mathbb{Z}}\left(-\frac{2\lambda N^2}{3^7\pi^2\beta^4}\left[1+\left(\frac{\lambda}{4\pi^2N}\right)^2(\theta+2\pi k)^2 \right]^{-3}\right)
\ee
A schematic plot of the branches is given in figure \ref{three_branches}. The branches cross many times as $\theta$ climbs (though as we will argue below, transitions are exponentially suppressed if one is not too far from the minimum on one's branch.) The energy of the true ground state as a function of $\theta$  is shown on the left hand side of figure \ref{vacstr}.  This structure is also consistent with lattice results \cite{Giusti:2007tu}: the potential is periodic, but rises quadratically from minima at $\theta = 2\pi n$, $n\in \Z$ with sharply kinked maxima at $\theta = (2n + 1)\pi $ joining the quadratic wells.

Again, we note that while $x$ is a periodic variable, with periodicity $\lambda/2\pi N$, we will often use the term "large $x$".  By this we are referring to the infinite cover of the ciricle on which $x$ lives, and in particular we mean that we have followed a given family of metastable states adiabatically from zero energy as $x$ increases in this infinite cover.

We can also study the theory at large $\lambda/N$, where $e^{\phi_0} \gg 1$, by working in the M theory frame.  We find that a branch can remain the true vacuum well away from the quadratic regime, and the energy rapidly reaches the saturation point before joining onto the next branch. This is shown in the figure on the right in figure \ref{vacstr}.

\subsection{Glueball masses}

We can estimate the scale of the glueball masses, and thus of the mass gap of the theory, by studying the spectrum of excitations of a metric perturbation \cite{Witten:1998zw,Gross:1998gk,Csaki:1998qr,deMelloKoch:1998qs}.

We find it easier to work in the M theory frame, with the background metric (\ref{eq:Mdecouple}).\  Consider a gravitational wave propagating along the $y_3$ direction.  The linearized wave equation for a metric perturbation $\delta h_{y_1y_2}=h(r)\exp[-i\omega t+ik_3 y_3+im\chi/\beta]$ is:footnote{Note that although we work with the metric in M theory, $\omega$ is conjugate to the time $t$ in type IIA and in the conjugate gauge theory.}
\be
h''+\left(\frac{3}{u f}-\frac{1}{u} \right)h'+\left( \frac{1-3/f}{u^2}+\frac{\lambda\beta(\omega^2-k_3^2)}{2u^3f}-\frac{n^2\lambda H_0}{2u^3f^2\beta}\right)h=0\label{eq:gluelineq}
\ee
We further demand regularity at the tip $u = u_0$ and and normalizability at infinity, leading to a quantized spectrum for $\omega$. 
First, let us analyze the case of zero KK charge, $m=0$. Rewriting in terms of $z=u_0/u$ we find
\be
h''(z)-\frac{3z^2}{1-z^3}h'(z)+\left(\frac{9\Omega^2}{4 z(1-z^3)}-\frac{2+z^3}{z^2(1-z^3)} \right) h(z)=0,\label{noscalewave}
\ee
where
\be
\Omega^2=\frac{2\lambda\beta}{9u_0}(\omega^2-k^2)=\beta^2(1+x^2)(\omega^2-k^2)
\ee
From this scaling alone, we know there is a spectrum of glueballs at energies
\be
\omega_g = \frac{1}{\beta}\sqrt{\frac{1}{1+x^2}}\times\Omega_g\label{glueballs}
\ee 
where $\Omega_g$ is an order one number. A numerical study of the normal modes of (\ref{noscalewave}) gives the first glueball at $\Omega_g\approx 1.57$ and they approximately follow a linear trajectory, $\Omega\approx 1.6+0.9 k$ where $k=0,1,2,\ldots$.
\begin{figure}[ht!]
\begin{center}
\includegraphics[scale=1]{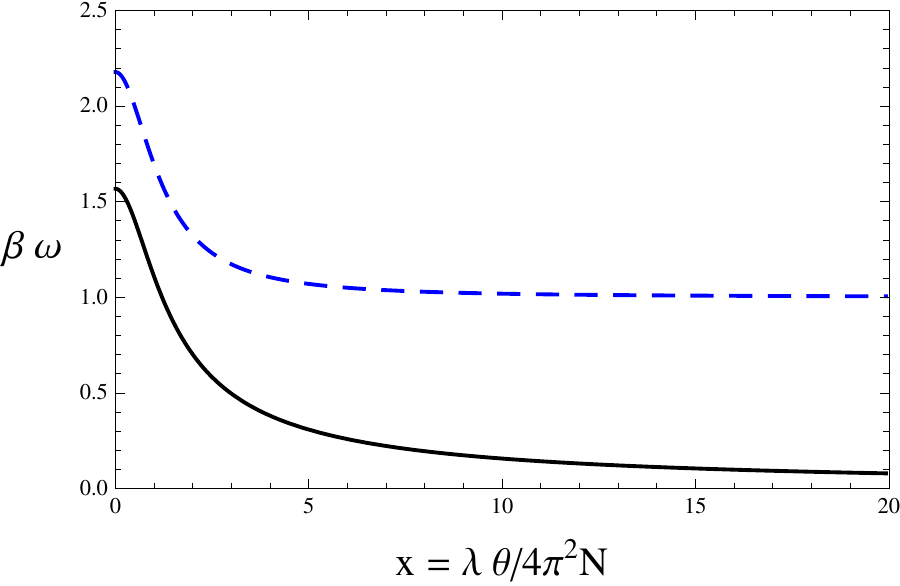}\end{center}
\vspace{-.5cm}
\caption{\label{spectra}A plot of the lightest glueball and lightest KK mode mass as a function of $x$. The lower solid  curve is the lightest glueball mass, and the upper curve is the lightest KK mode ($n=1$.) Note that the glueball mass becomes parametrically smaller than the KK scale as we increase $x$.}
\end{figure}

It is clear from (\ref{glueballs}) that the glueball masses become light at large $\theta$.  This is consistent with the fact that the ``end'' $u_0$ of the radial direction, which is a measure of the confinement scale, becomes small at large $\theta$.  It is also worth noting that the masses become parametrically smaller than the KK scale at which the field theory becomes five-dimensional. Note that as soon as $n\neq 0$, we can not simply scale all parameters out of the wave equation. However, we can analyze the system at  large $x$, where the geometry is nearly that of the deconfined phase. In the deconfined phase we know the KK spectrum is $\omega_m=\frac{m}{\beta}$, which implies that at large $x$ there is an explicit separation of scales between the glueball and KK spectra,
\be
\omega_{glue}=\frac{\Omega_g}{\beta x}+\Op(1/x^3),~\omega_{KK}=\frac{m}{\beta}+\Op(1/x^2).
\ee

\subsection{Domain walls and the field theory landscape}

At fixed $\theta$, there is a stable vacuum and a landscape of metastable vacua.  The interpolating fields are nonperturbative excitations of the field theory (either of the low-energy 4d field theory or of the 5d theory controlling the UV fixed point -- this depends in part on the energy of the potential barriers.)  Beginning in a stable vacuum at, say, $\theta = 0$, and adiabatically changing $\theta$, the vacuum becomes metastable.  One expects a domain wall which interpolates between adjacent branches.

For the present theory at small $\theta$, one such domain wall is a D6-brane wrapping the $S^4$ and sitting at $u = u_0$ \cite{Witten:1998uka}. We call this the ``thin'' domain wall.  We will identify it and estimate its tension as a function of $x$.  There is a related ``thick'' domain wall configuration consisting of a D6-brane which wraps an $S^3 \subset S^4$ at a fixed value of the direction transverse to the domain wall, and which sweeps out the entire $S^4$ as one crosses the domain wall.  This is directly analogous to the domain wall found in \cite{Kachru:2002gs}.  

Given such domain walls, there are nonperturbative instability for hopping between branches, arising from the nucleation of a critical bubble as in \cite{Coleman:1977py}.  We will discuss the potential instabilities in \S4; we will find that at large $x$, the ``thick'' domain wall nucleates more rapidly at large $x$ than the ``thin'' wall.

\subsubsection{Thin domain walls}

We denote by ``thin'' domain walls the domain walls which come from wrapping a D6-brane around the $S^4$.  This domain wall will not be static.  In the bulk, the force on the domain wall can be extracted from the D-brane action directly.  In the dual theory, the force per unit area on the domain wall should be equal to the difference in energy density between the two metastable vacua.  We will find that these two expressions match in the limit that the RR magnetic 2-form flux is large, as gauge-gravity duality demands.

We write the spatial directions of the 4d theory in spherical coordinates $\rho,\Omega_2$, and consider a D6-brane wrapping the $S^4$, spherically symmetric in the field theory directions, and moving in $\rho, u$. This describes a family of spherical domain walls.  In the deconfined phase, at weak coupling, the D6 is codimension 2 along the D4 worldvolume directions, and so it should feel an attractive force towards the D4-branes. The strong coupling analog is that the D6 is attracted to the point $u = u_0$, which we will find to be the case.

In the limit of large RR 2-form flux, the D6-brane action can be treated in the probe approximation: The 2-form flux jumps across the wall; the probe approximation amounts to ignoring this backreaction, which is a good approximation when the difference in flux is small compared to the total flux.  The DBI action is
\be
S_{DBI,6}=-\mu_6\int d^{7}\xi e^{-\phi}\sqrt{-\det G_{\alpha\beta}}\label{eq:dbisix}
\ee
where $G_{\alpha\beta}=X^a_{,\alpha}X^b_{,\beta}g_{ab}$ is the pullback of the spacetime string frame metric to the D6 worldvolume, $\xi^\alpha$ are the worldvolume coordinates, and $X^a$ are the spacetime coordinates.  The Chern-Simons (CS) term is simply the integrated pullback of the RR potential
\be
S_{CS,6}=+\mu_6\int d^{7}\xi\left[ X^{a_1}_{,\alpha_1}\ldots X^{a_7}_{,\alpha_7} C^{(7)}_{a_1 \ldots a_7} \right]\label{eq:cssix}
\ee
From $dC^{(7)}= dC^{(1)}$, we find, in a specific gauge:
\be
C^{(7)}=\ap^{7/2}\frac{16\pi \lambda^2Nx\rho^3 }{3^6\beta^4(1+x^2)^4}~dt\wedge \epsilon^{(2)}\wedge \epsilon^{(4)}_\Omega
\ee
If we demand that $\rho, u, t$ depend only on the worldsheet coordinate $\xi^0$, and fix to static gauge $t = \xi^0$, then the sum of  (\ref{eq:dbisix},\ref{eq:cssix}) reduces to the Lagrangian of a particle in two dimensions:
\be
S_{D6} = \int dt\left[-\frac{N H_0 u \rho^2}{6 \pi^2}\sqrt{\frac{2u^3}{\beta\lambda}-\frac{\dot{u}^2}{f}-\frac{2u^3\dot{\rho}^2}{\beta\lambda}}+\frac{8N x \lambda^2\rho^3}{3^7\pi^2\beta^4(1+x^2)^4} \right]\ .
\ee
The potential as a function of $u$ has no stationary points, and a minimum at $u=u_0$. At this point, the action is:
\be
S = \int dt \frac{4N \lambda^2}{3^6\pi^2}\left[-\frac{\rho^2\sqrt{1-\dot{\rho}^2}}{\beta^3(1+x^2)^{7/2}} +\frac{2 x\rho^3}{3 \beta^4(1+x^2)^4}\right]\ .
\ee

The energy $U$ for an initially static bubble of new vacuum can be interpreted as coming from two parts - the tension $\tau$ of the domain wall, and the energy difference $\Delta\CE$ between the two branches inside the bubble:
\be
%S = - \int dt U(\rho)\ ; 
U  = \tau \times 4 \pi \rho^2 + \Delta \CE \times\frac{4\pi\rho^3}{3}\ ,\label{eq:tensiontw}
\ee
where
\be
\tau=\frac{\lambda^2 N}{3^6\pi^3\beta^3(1+x^2)^{7/2}}\ ,\label{eq:energytw}
\ee
and
\be\label{eq:branchdiff}
\Delta\CE = - \frac{2\lambda^2 N x}{3^6\pi^3\beta^4(1+x^2)^4} \ .
\ee
The tension scales with $N$, as found in \cite{Witten:1998uka}.  As $x$ increases, both the tension and the energy difference decrease, consistent with our observation that the dynamical scale of the gauge theory decreases at large $x$.  

There are two especially interesting initially static bubbles. First, there is a single extremum of the action when $\dot{\rho} = 0$, corresponding to an unstable critical bubble (sphaleron), at
\be
\rho_{un} = \frac{\sqrt{1+x^2}}{x}\beta\ .\label{eq:critsize}
\ee
Secondly, there are zero energy bubbles which can nucleate from the vacuum via a semiclassical instanton,  with zero kinetic energy at the point of nucleation  \cite{Coleman:1977py}.  The radius of these bubbles (which is just the radius of the Euclidean instanton) can be found by setting the energy to zero:
\be
	\rho_{nuc} = \frac{3}{2} \frac{\sqrt{1 + x^2}}{x} \beta = -\frac{3\tau}{\Delta\CE}\ .\label{eq:nucsize}
\ee
The fact this is larger than the radius of the critical bubble means that once a bubble of size $\rho_{nuc}$ nucleates it will expand as more and more of the spacetime makes the transition to the lower branch of the potential $E(\theta)$.

At large $N$ and fixed $\lambda$, the energy difference $\Delta\CE$ agrees with what we would predict from (\ref{thetapotential}).  Using the fact that that the difference between adjacent branches is equal to the difference between energies along a single branch as $\theta$ is shifted by $2\pi$, we find that to leading order in $1/N$,
\be
E(\theta-2 \pi)/V -E(\theta)/V \sim  -\frac{\lambda}{2\pi N} \frac{\p E(\theta(x))}{\p x} = 
\Delta \CE.\label{energydiff}
\ee

\subsubsection{Thick domain wall}

A different configuration of wrapped D6-branes arises from a second class of domain wall, which is in close analogy to the domain wall discussed in \cite{Kachru:2002gs}.  We again work in the probe approximation for the D6-brane, valid when the RR 2-form flux is large.  Consider a D6-brane which fills the 4d field theory directions $y_i$ and wraps one of a family of $S^3$s sitting inside the $S^4$.  Write the $S^4$ metric as
\be
d\varphi^2+\sin^2\varphi d\Omega_3^2,\label{eq:Sfourfoliation}
\ee
where $\varphi$ parametrizes the family of $S^3$s that the D6-brane wraps, and will be a general function of $t,y_i$.  The 7-form potential is:
\be
C^{(7)}=\frac{2^6\pi \ap^{7/2}\lambda^2 N}{3^6\beta^4(1+x^2)^4}\sin^4(\varphi/2) (2+\cos\varphi)dt\wedge dy_1\wedge dy_2\wedge dy_3 \wedge d \epsilon^{(3)}_\Omega\ ,\label{eq:sevenformthick}
\ee
where we have fixed a gauge such that the total action vanishes when the brane slips off the pole and vanishes at $\varphi=0$.  The full DBI+CS action for the D6-brane as a function of $\varphi$ is
\begin{eqnarray}
S_{D6} & = & \int dtd^3 y \left[-\frac{H_0 N u^{3/2}\sin^3\varphi}{16\pi^3\beta\lambda}\sqrt{u^3-\frac{\beta\lambda (\p u)^2}{2f}-\frac{u^2\beta\lambda(\p\varphi)^2}{2\sin^2\varphi}} + \right.\nonumber\\
& & \ \ \ \ \ \left.+\frac{2\lambda^2 N x\sin^4(\varphi/2)(2+\cos\varphi)}{3^6\pi^3\beta^4(1+x^2)^4}\right]\label{eq:thickdomainmode}
\end{eqnarray}
We again find that the D6 wants to sit at $u = u_0$. 

\begin{figure}[ht!]
\begin{center}
\includegraphics[height=3in]{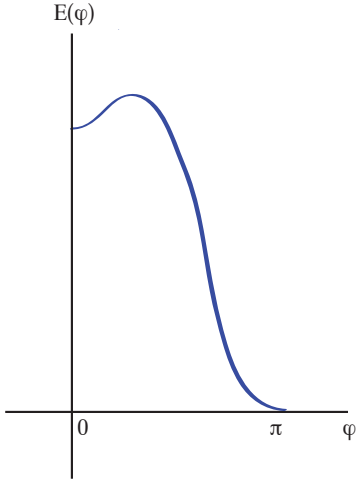}\end{center}
\vspace{-.5cm}
\caption{\label{branepot}A plot of the potential energy density of a D6-brane at $u = u_0$ wrapping the $S^3\subset S^4$ labeled by $\varphi$.}
\end{figure}

Now imagine that we adiabatically move $\varphi$ from $0$ to $\phi$.  At finite $\varphi$, the 8-form electric flux $dC^{(7)}$, or the dual 2-form magnetic flux, jumps across the D6-brane. At $\varphi = \pi$, the D6-brane disappears and the 8-form electric flux has shifted by one unit -- we have interpolated between adjacent branches of the field theory.  Thus, the position $\varphi$ is a field theory mode interpolating between adjacent branches at fixed $\theta$. A sketch of the potential energy for this mode (found by setting $\p u = \p \varphi = 0$ in (]\ref{eq:thickdomainmode}) can be seen in Figure \ref{branepot}.  A sense of the shape of the potential can be given by the value of $\phi$ at the maximum of the potential:
\be
\varphi_{max}=\mathrm{arccot}(x) \label{eq:potmax}\ ,
\ee
and by the energy density at this maximum with respect to the higher-energy branch at $\varphi = 0$:
\be
\CE_{unstable}=\frac{\lambda^2 N}{2\times3^6\pi^3\beta^4}\frac{2x^2+1-2x\sqrt{1+x^2}}{(1+x^2)^{9/2}}\approx \frac{\lambda^2 N}{2^3 3^6\pi^3\beta^4 x^{11}}+\Op(1/x^{12})
\ee
As $x$ gets larger and larger, $\varphi_{max}$ is pushed towards $\varphi = 0$; the barrier gets smaller and smaller and close and closer to $\varphi = 0$, indicating that an instability is developing.  We will discuss this further in the next section. Note also that the energy difference between the vacua at $\varphi = 0$ and $\varphi = \pi$ is still given by (\ref{eq:branchdiff}).  At large $x$, the height of the barriers $\sim 1/x^{11}$ are much smaller than the energy difference between the vacua $\Delta \CE \sim 1/x^7$. 

\begin{figure}[ht!]
\begin{center}
\includegraphics[height=4in]{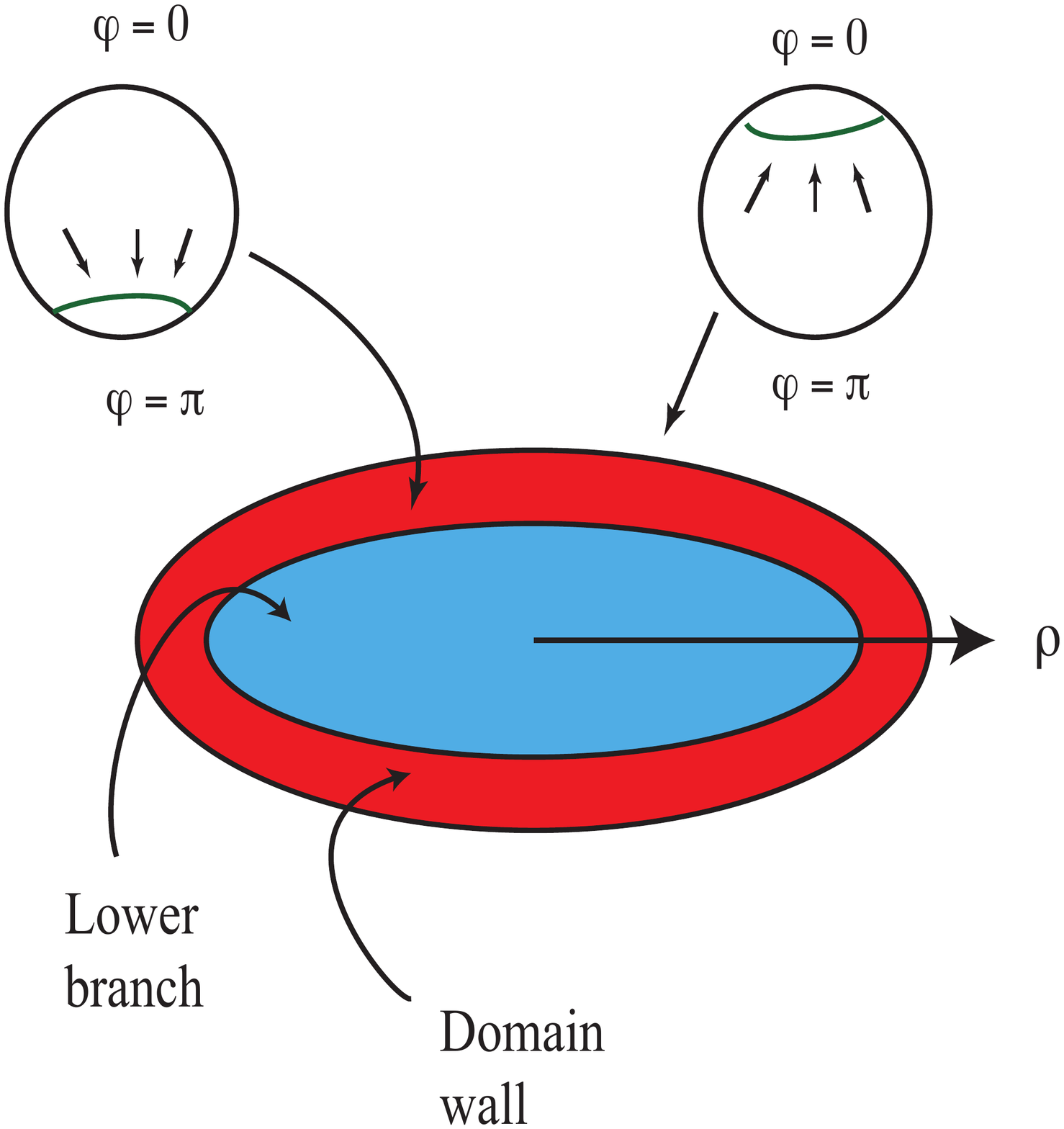}\end{center}
\vspace{-.5cm}
\caption{\label{domainbubble} A schematic picture of a bubble of thick domain wall. A D6-brane slips from the north pole towards the south pole as one moves into the interior of the bubble.  Note that upon initial nucleation of the bubble, $\varphi < \pi$; the D6-brane continues to slip off the south pole during the subsequent classical evolution of the bubble.
}
\end{figure}

It is clear that we can describe a domain wall for which $\varphi$ interpolates between $0$ and $\pi$ on each side of the wall; and we can write a spherical bubble in which $\varphi$ varies radially from $\varphi = 0$ outside of the bubble to $\varphi = \pi$ inside the bubble. This is demonstrated schematically in figure \ref{domainbubble}.  One can find that generically such ``thick'' walls have larger tension, due to the cost of stretching the D6 out in the radial direction. Despite this, we will find below that at large $x$, the instanton for nucleating this bubble has much lower action than the instanton for nucleating the ``thin-wall'' bubble discussed above, because the instanton can be made very small by being very localized on the $S^4$. 

\section{Limits on the range of axions}

One motivation of this work was to understand whether a given branch of metastable vacua would cease to be even metastable. As motivation, we can consider another theory which exhibits axion monodromy, Yang-Mills coupled to fundamental fermions with slightly unequal masses.  The energy as a function of $\theta$ can be studied using chiral lagrangian techniques, following \cite{Witten:1980sp}.  Ref. \cite{Dubovsky:2010je}\ showed explicity that a given branch of metastable states will, as $\theta$ increases, merge with a branch of saddle points and disappear as any kind of stationary point in the effective potential for the mesons.  In this section we will argue that the metastable branches in our strongly-coupled gauge theory will also end, via one of several candidate mechanisms.  Which mechanism dominates will depend on the details of how one takes the large-N, large-$\lambda$ limit.

The first class of mechanisms is perturbative in the bulk string coupling.  As $\theta$ increases, the $u-\chi$ directions in the IR look more and more like a long, thin, cylinder with slowly shrinking radius.  Tachyonic winding strings or Casimir forces will cause the cylinder to pinch off as the size of the cylinder locally becomes of order string scale; such tachyons are known to destabilize the "deconfined" solution, following \cite{Horowitz:2005vp}. It is possible that one could continue to increase the RR 2-form flux while the tip of the $u-\chi$ throat ceases to recede to the IR.  We have no evidence for such a class of solutions, and we conjecture that they do not exist; if this conjecture is true, it means that stable vacua cease to exist.

The second class of mechanisms is nonperturbative in nature: ``thin'' or ``thick'' domain walls can nucleate, driving the solution to a lower branch.  We expect the higher-energy branch to move from being metastable to unstable roughly when the action for bubble nucleation becomes of order $1$.  
Furthermore, we will find that at large $\theta$, the dominant mode will be the ``thick'' domain walls described in \S3.3.2.  The corresponding field theory mode has a potential barrier between the two branches . The height decreases as a function of $x$, making clear that the branch will become unstable and cease to exist for large enough $\theta$.  

To aid the reader, we first list all of our results: the instabilities of our confined geometry, both perturbative and nonperturbative, and the bounds they put on how far out on a single vacua we can traverse in terms of $x=\lambda \theta/4\pi^2 N$.

\begin{table}[htdp]
\caption{default}
\begin{center}
\begin{tabular}{|c|c|c|}
\hline
Instability & Type & Bound on stable range\\
\hline\hline
Winding strings & Perturbative & $x < \lambda^{1/3}.$\\\hline
Casimir energy & Perturbative & $x^7 \ll N \lambda^{1/2}$\\\hline
Thin bubble nucleation & Nonperturbative & $x^7 \ll \lambda^2 N$\\\hline
Thick bubble nucleation & Nonperturbative & $x^{11}\ll \lambda^2 N$ \\\hline
\end{tabular}
\end{center}
\label{default}
\end{table}%

\subsection{Perturbative instabilities}

At large values of $\theta$, there is a regime for which the size of the $\chi$ circle is changing slowly as a function of proper distance along $u$, in a region of the geometry far from the tip at $u = u_0$:
\be
	\epsilon = \sqrt{g^{uu}}\frac{\p R_{\chi,proper}(u)}{\p u} \ll 1 \ .
\ee
where the proper size $R_{\chi}$ of the $\chi$ circle at fixed $u$ is, according to  (\ref{D4throat}):
\be\label{eq:properradius}
	R_{\chi,proper}(u) = \frac{\sqrt{\alpha'} f^{1/2} u^{3/4} \beta}{H_0^{1/4} (\lambda\beta)^{1/4}}\ ,
\ee
and 
\be
	\sqrt{g^{uu}} = \frac{f^{1/2} u^{3/4}}{\sqrt{\alpha'} (\lambda\beta)^{1/4} H_0^{1/4}}\ .
\ee
We will be interested in the regime for which $\eps \ll 1$ far from $u = u_0$, in which case we can approximate $f , H_0 = 1$.  In this limit,
\be
	\epsilon = \sqrt{g^{uu}}\frac{\p R_{\tau,proper}(u)}{\p u} \sim \left(\frac{u \beta}{\lambda}\right)^{1/2}
\ee
and the slope $\epsilon \ll 1$ if 
\be
	u \ll \frac{\lambda}{\beta}
\ee
Now recall that $u_0 = \frac{\lambda}{\beta} \cos^2\gamma$.  Thus, we only have a nearly-flat cylindrical region if $\cos\gamma \ll 1$, that is if $x \gg 1$. 

\subsubsection{String winding modes}

Since the spacetime fermions have antiperiodic boundary conditions around the circle $\tau$, we expect that the $(mass)^2$ of strings winding this circle to have a tachyonic mass shift of order $- m_s^2$.  Thus, a winding tachyon will develop when the proper size of the $\chi$ circle at fixed $u$ is of order the string scale:
\be
	R_{proper,\chi} = m_s \frac{(u \beta)^{3/4}}{\lambda^{1/4}} \sim \sqrt{\alpha'}
\ee
which will occur when
\be\label{eq:windingcondition}
	u \sim \frac{\lambda^{1/3}}{\beta} = \lambda^{-2/3} \frac{\lambda}{\beta}
\ee
Now since $u > u_0 = (\lambda/\beta) \cos^2\gamma$, (\ref{eq:windingcondition}) cannot be met unless $\cos^2\gamma < \lambda^{-2/3}$, which for $\lambda \gg 1$ means that
\be
	x \gg \lambda^{1/3}\ ;\ \ \ \theta \gg N/\lambda^{2/3}\label{eq:windingcond}
\ee

 In this limit, we expect that the winding tachyon will condense and cause the solution to ``cap off'' the $u-\chi$ cylinder, much as in \cite{Adams:2005rb,Kruczenski:2005pj}.  As we argued above, we conjecture that the branch simply ceases to exist as a metastable confuguration of the field theory in this regime.

\subsubsection{Casimir energy}

In the presence of SUSY-breaking fermion boundary conditions about $\tau$, we expect that the Casimir energy will cause the $\chi$ circle to want to shrink and pinch off.  Again, we will focus on the region where $\epsilon \ll 1$ (which only exists for large $x$); at large values of $u$, the circle is stabilized by the asymptotic boundary conditions.

For small values of $u$, the finite radius of the $u-\chi$ cigar is supported by the RR two-form.  Thus we will compare the energy density due to the 2-form to the energy density due to the Casimir energy, to see when the latter dominates.  We first wish to determine whether the Casimir energy should be computed in 6 dimensions $(t, \vec{y}, u, \tau)$ or ten dimensions (by including the $S^4$).  The latter will be correct if the proper radius of the $S^4$ is large compared to $R_{proper,\tau}$ in the regime $\epsilon \ll 1$.  The proper radius of the $S^4$ can be read off of the metric (\ref{D4throat}), and is:
\be
	R_{S^4} \sim \sqrt{\alpha'} H_0(u) (\lambda \beta u)^{1/4}
\ee
Thus
\be
	\frac{R_{proper,\tau}}{R_{S^4}} \sim \left(\frac{u \beta}{\lambda}\right)^{1/2}
\ee
This is small (and so the Casimir energy is computed in 10 dimensions) precisely when
$u < \lambda/\beta$, which is when the $u-\chi$ directions approximate a straight cylinder.

In this regime, the contribution of the two-form to the action in string frame is:
\be
	\frac{1}{(\alpha')^3}\int d^{10} x \sqrt{g} g^{uu} g^{\tau\tau} (F_{u\tau})^2
\ee
Now $g^{uu} g^{\tau\tau} \sim (\alpha')^{-2}$.  Thus, we can write
\be
	C_{\tau} = \frac{1}{R_{11}} \frac{1}{H_0} \cot\gamma + {\rm constant}\ ,
\ee
and so
\be
	F_{\tau u} \sim \frac{3 u_0^3}{R_{11} u^4} \frac{\sin\gamma\cos\gamma}{H_0^2}
\ee
Thus the potential due to the 2-form (where we approximate $H_0 \sim 1$) is
\be
	V_{2f} \sim \frac{ 9 u_0^6 \sin^2\gamma\cos^2\gamma}{(\alpha')^5 R_{11}^4 u^8}
\ee
The Casimir energy density is
\be
	V_{cas} \sim \frac{1}{R_{\tau,proper}^{10}} \sim \frac{\lambda^{5/2} \beta^{5/2}}{(\alpha')^5 u^{15/2} \beta^{10}}\ .
\ee
The ratio is:
\be
	\frac{V_{cas}}{V_{2f}} \sim \frac{\lambda^{5/2}}{(\beta u)^{15/2}} \frac{u^8 R_{11}^2}{u_0^6 \sin^2\gamma\cos^2\gamma}
\ee
Using $u_0 \sim (\lambda/\beta) \cos^2\gamma$ and $R_{11} = \lambda \beta/N$,
\be
	\frac{V_{cas}}{V_{2f}} \sim \frac{1}{N^2\lambda} \left(\frac{u\beta}{\lambda}\right)\frac{1}{\sin^2\gamma \cos^{14}\gamma}\ .
\ee
Since $\cos\gamma \sim 1/x$ at large $x$, the Casimir energy is sub-dominant so long as
\be
	x^{14} \ll N^2\lambda \left(\frac{\lambda}{\beta u}\right)^{1/2}\ .
\ee
The right hand side is minimized when $u$ is largest. Recall that we are demanding that $\epsilon \ll 1$, which occurs when $u < \lambda/\beta$.  In this regime, the condition for the solution to be stable against the Casimir force is
\be
	x^7 \ll N \lambda^{1/2}%\ ; \ \ \ \theta \gg N/\lambda^{1/2}
\ee
which is a less stringent requirement than (\ref{eq:windingcond}).

\subsection{Nonperturbative instabilities}

In addition to the perturbative instabilities described above, bubbles of a lower branch can nucleate, bounded by one of the domain walls described in \S3.3.  We will find that for large $x$, nucleation of the ``thick domain wall'' is the dominant transition.

\subsubsection{Nucleation of the thin domain wall}

For large $\theta$, we can compute the action of the ``thin'' D6-brane in the bulk, using the probe limit.  The critical bubble size (\ref{eq:nucsize}) is $\rho_{nuc} \sim 3 \beta/2$ at large $x$.  In the field theory, this means the nucleation rate should be computed in the 5d theory.  We are computing it in the dual 10d theory, in the limit $x \gg 1$, where we can use the probe limit to describe the instanton action.   However, we found in \S3.3.1\ that the action and critical size of a bubble at fixed $\rho$ is identical to the expression in the 4d thin wall approximation, for a bubble with tension $\tau$ (\ref{eq:tensiontw}), enclosing a region with energy lower by $\Delta\CE$  (\ref{eq:energytw}).  The resulting decay probability will be proportional to
\be
	\Gamma \sim e^{-S_{inst}}\ , S_{inst} = \frac{27\pi^2}{2} \frac{\tau^4}{\Delta\CE^3} = \frac{\lambda^2 N}{8 3^6 \pi^3 \left(1 + x^2\right)^2 x^3} \sim \frac{\lambda^2 N}{x^7}\ .\label{eq:instacttw}
\ee
The action therefore decreases at large $\theta$, indicating that eventually the metastable vacua at large $\theta$ will cease to be stable. A similar phenomena was found for softly broken SUSY theories in \cite{Shifman:1998if}.  In the present example, as we will now find, at large $x$ nucleation of the ``thick'' domain walls dominates in this regime, and the large-$\theta$ branch will become unstable even sooner that would be indicated by the decreasing action of these thin domain walls.

\subsubsection{Thick domain walls}

The second channel for discharging RR 2-form flux, and thus changing branches, is via the nucleation of a ``thick domain wall''.  More precisely, the bubble can be constructed as shown in Fig. \ref{domainbubble}.  As we approach the bubble from infinity in the radial direction, a D6-brane appears and wraps the $S^3$ near the north pole $\varphi = 0$ of the sphere (\ref{eq:Sfourfoliation}), while staying at $u = u_0$ and filling out the $t, y_i$ directions.  The RR 2-form flux seen by an observer at the north pole is now down by one unit from the RR 2-form flux seen by an observer at the south pole.  As we continue towards the origin of the bubble, the position $\varphi$ of the D6-brane increases.  We have found numerically that when $x \gg 1$, $\varphi$ has not yet reached $\pi$ at the center of the bubble.  As the bubble expands and evolves, the D6-brane will eventually slip off the south pole at the center.  The region where $\varphi = \pi$ will grow, as more and more of the spacetime lies in the lower branch of the potential.

Again, the nucleation of this bubble is well-described by constructing the $SO(4)$-invariant instanton.  
We work in spherical coordinates $\rho, \Omega'_3$ on $\RR^4$.  The seven-form (\ref{eq:sevenformthick}) in these coordinates is: 
\be
C^{(7)}=\frac{2^6\pi \ap^{7/2}\lambda^2 N}{3^6\beta^4(1+x^2)^4}\sin^4(\varphi/2) (2+\cos\varphi)\rho^3 d\rho\wedge d\epsilon^{(3)}_{\Omega'} \wedge d \epsilon^{(3)}_\Omega
\ee
The action is minimized if the D6-brane sits at constant $u = u_0$.   The resulting brane configuration describes a curve in $(\varphi, \rho)$.  The action for this curve, parametrized by $\xi^0$, is:
\be
S=\frac{\lambda^2N}{2\times3^6\pi\beta^4(1+x^2)^4}\int d\xi^0 \times \rho^3\left[ \sin^3\varphi\sqrt{4\rho'^2+9(1+x^2)\beta^2\varphi'^2}-8 x\rho'(2+\cos\varphi)\sin^4(\varphi/2)\right]\ ,
\ee
where primes denote derivatives with respect to $\xi^0$.

\begin{figure}[ht!]
\begin{center}
\includegraphics[height=3in]{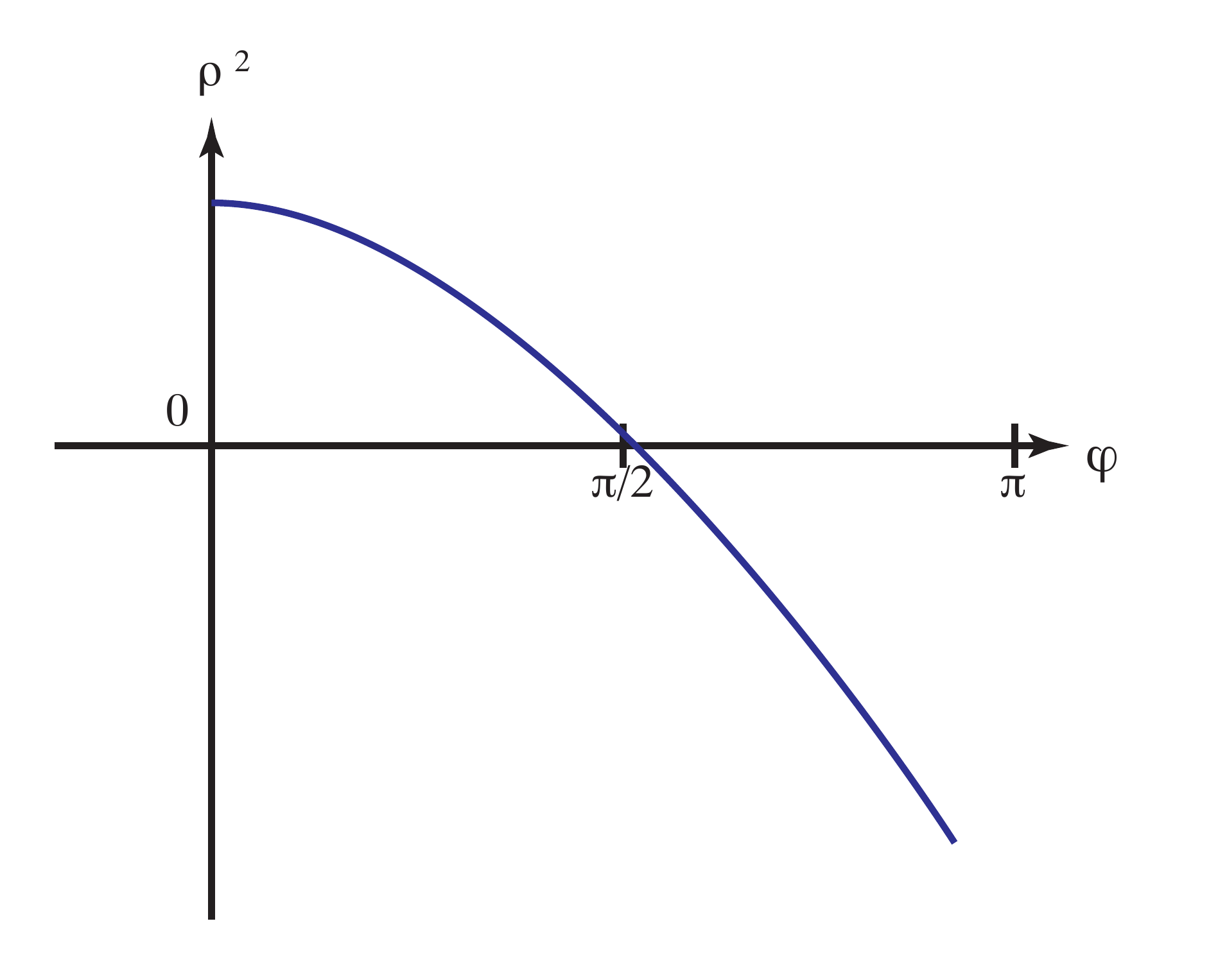}\end{center}
\vspace{-.5cm}
\caption{\label{braneprofile} The profile $\varphi(\rho^2)$ for $x \sim 5$.  For $\rho^2 > 0$, this describes the Euclidean instanton.  The continuation to $\rho^2 = - t^2 + \vec{y}^2 < 0$ determines part of the Lorentzian evolution.
%Note that as $x$ increases the $\rho^2>0$ component of the instanton is located only near the north pole at $\varphi=0$.
}
\end{figure}

Figure 7 shows the profile $\varphi(\rho^2)$ for $x > 1$.  As $x$ increases, we find that the solution in the Euclidean regime becomes more and more concentrated near $\varphi = 0$.  The point is that the instanton describes the classically forbidden regime at energies equal to the higher-energy vacuum.  This regime shrinks at large $x$.  In Figure 7 we have continued the solution to $\rho^2 < 0$, in order to describe the Lorentzian solution.  In Lorentzian signature  $\rho^2 = - t^2 +\vec{y}^2$. $\varphi$ remains a function of $\rho$ only.  The initial bubble corresponds to a slice of the Euclidean solution through the origin.  This slice is taken to be the $t = 0$ initial surface in Lorentzian signature.  As seen in Figure (\ref{braneprofile}), $\rho^2 = 0$ is described by the (forward) light cone emanating from the origin.  The surfaces of constant $\rho^2 > 0$ form timelike hyperboloids outside of this light cone; the surfaces of constant $\rho^2 < 0$ form spacelike hyperboloids inside of the light cone.
The value of $\varphi$ at $\rho = 0$ is on the right side of the barrier shown in Fig. (\ref{branepot}).  Thus, in Lorentzian signature, $\varphi$ will continue to roll to $\pi$ as the bubble evolves and expands.  At this point we do not know what happens when the brane disappears.  The kinetic energy could be dumped into radiation.  Alternatively the system could be driven classically to make a further transition to the next lower branch, if the potential barrier height for that transition is low enough.\footnote{We would like to thank Matt Kleban for discussions of this point.}
 
In order to estimate the rate of nucleation of "thick wall" bubbles, we can extract the scaling of the instanton action with $x \gg 1$ by defining
\be
\varphi =  \Phi/x,~\rho = \beta r,~\xi^0 = \beta \zeta 
\ee
The action in these variables is:
\be
S=\frac{\lambda^2 N}{ 3^5\times4\pi x^{11}}\int d\zeta \left[2r^3\Phi^3\sqrt{4r'^2+9\Phi'^2} -3r^3\Phi^4 r'\right]+\Op(1/x^{12})
\ee
where primes in this expression denote derivatives with respect to $\zeta$. Since the tunneling rate scales as $\Gamma \sim e^{-S}$, we find that the rate becomes large as
\be
\lambda^2N \gg x^{11}.
\ee
It is clear that in this regime, the potential barrier seen in Fig.~\ref{branepot} has vanished and the large-$x$ branch has ended.  

\begin{figure}[ht!]
\begin{center}
\includegraphics[height = 3in]{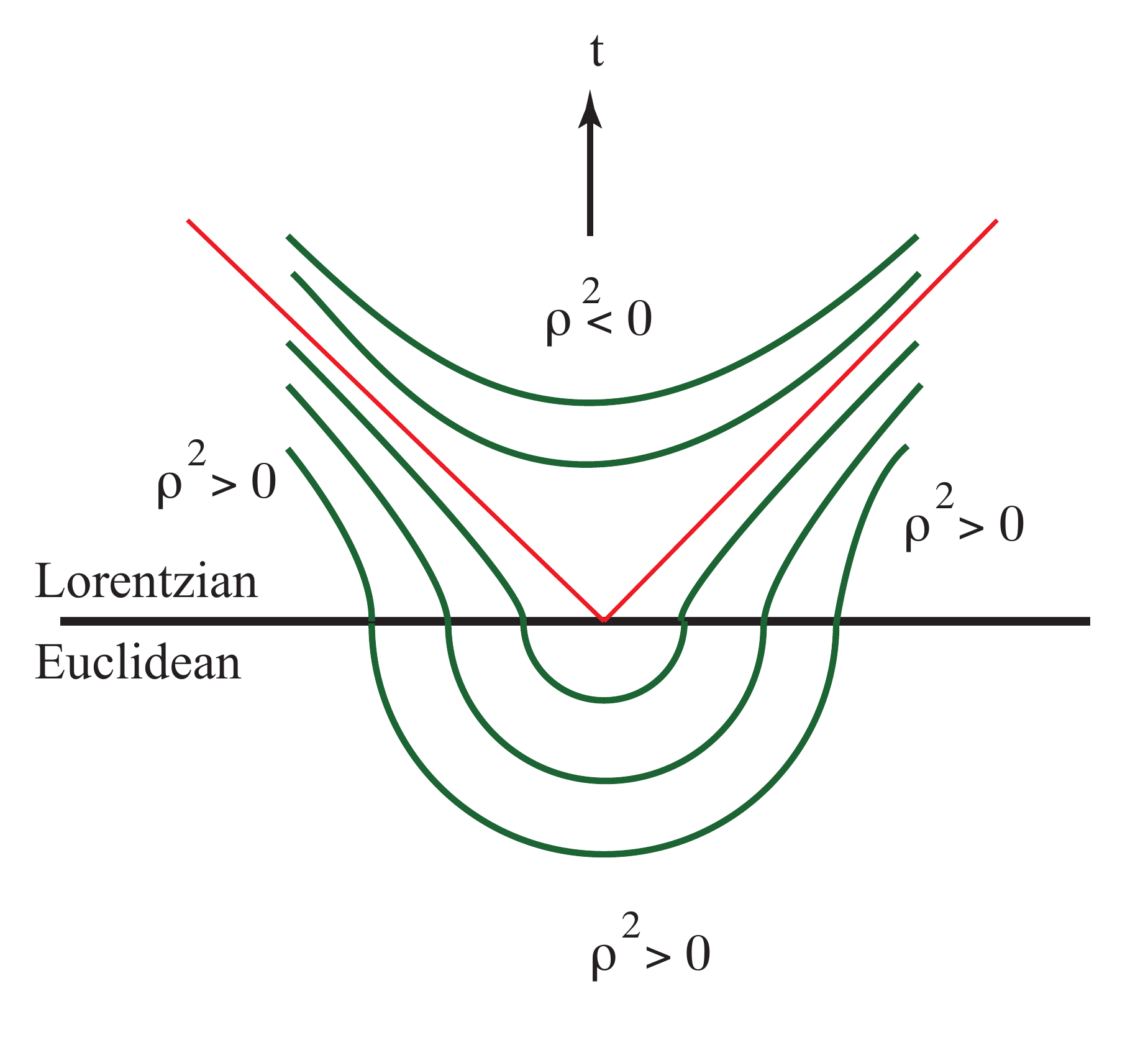}\end{center}
\vspace{-.5cm}
\caption{\label{bubblenuc} Lines of constant $\rho^2$ and therefore $\varphi(\rho^2)$.  Below the line is a hemispherical slice of the $SO(4)$-invariant Euclidean instanton. Above the line, we see the bubble expand and $\varphi$ continue to evolve inside the forward light cone (the lines emanating from the origin at 45 degree angles.)}
\end{figure}

\section{Dynamical axions}

Up until now, we have treated $\theta$ as a field theory parameter.  However, we expect the UV completion of our theory to be some form of string compactification, in which $\theta$ is promoted to a pseudoscalar axion.  This is the context in which the phenomenological applications described in the Introduction should be studied.   The full compactification, with the attendant moduli stabilization mechanism, will in general be quite complicated.  We can model its effects after the fashion of \cite{Randall:1999vf,Verlinde:1999fy}: by cutting off the geometry at some $u = u_{UV}$ and allowing the formerly "non-normalizable" bulk modes, which describe the field theory couplings, to fluctuate.

In this section we will discuss two phenomena which can arise when the axion becomes dynamical: axion-driven "monodromy" inflation, and axion strings which bound domain walls.  Before doing this, we will compute the axion decay constant and Planck scale induced by integrating out the field theory degrees of freedom.

We expect embedding our story into a full compactification of string or M theory to be complicated.
An immediate obstacle is that the size $\beta$ of the Scherk--Schwartz circle becomes a dynamical modulus, and must be stabilized. The field theory degrees of freedom contribute a runaway $1/\beta^4$ potential (\ref{thetapotential}); a contribution from the bulk must compete. The appearance of this light modulus is not accidental. If we were to consider "supersymmetric" boundary conditions around the circle, $\beta$ would be a scalar component of the axion supermultiplet. To stabilize this modulus without giving a mass to the axion an additional source of supersymmetry breaking is needed. This makes the construction of a concrete model much harder. Furthermore, it may be difficult to embed our field theory into a full IIA compactification from the start, with or without low-energy SUSY; for example, one must figure out how the $\chi$ circle continues, and ensure that $C_{\chi}$ appears as a modulus or pseudomodulus even though the geometry has no nontrivial one-cycle in the confined phase.

Our approach will be to see whether interesting lessons can be learnt  if one assumes that these issues can be resolved, and the $\beta$-modulus is stabilized without disturbing the axion potential. This is not totally ad hoc.  The very motivation for using the axion as an inflaton is that as a pseudo-Goldstone boson it is protected from perturbative corrections, including those which break supersymmetry.  In the context of our setup this means that axion potential does not receive perturbative corrections from the parts of the geometry where the $\beta$ cycle is non-contractable.
 
In the Introduction we discussed two distinct classes of applications of axion mondromy -- inflationary model building, and the coherent production of ultra-light axions. The specific setup discussed here is less useful for the latter application (though we hope that some general lessons will apply): the axion potential (\ref{thetapotential}) does not have an exponential sensitivity on the parameters of compactification that would most easily generate a low enough mass. This is related to the absence of separation of scales between  KK modes around the $\beta$-circle and  glueballs. It may be possible to obtain an ultra-light axion in this case by introducing additional source of warping.  Instead, we discuss here only possible applications for inflation.

\subsection{Embedding into a string/M theory compactification}

Upon embedding the theory into a full string compactification, the 4d couplings will fluctuate.  We will discuss the Planck scale, axion decay constant, and moduli potential here.  

If the field theory we describe here is a "sector" of the full string theory, it will be realized geometrically as a "throat" opening up inside some 6d string of 7d M-theory compactification.  We can model this by cutting off the geometries (\ref{D4throat}) or (\ref{eq:Mdecouple}) at some $u = u_{UV}$.  The coefficients for the kinetic terms of the 4d metric and axion, among others, will receive contributions from the bulk geometry and from the throat.   The throat contributions will be:
 \be
\delta M_{pl}^2\simeq {N^2 u_{UV}^2\over 24 \pi^2 \lambda },\;  \delta f_\phi\simeq {\lambda u_{UV}^2\over 192\pi^6}\ .\label{eq:throatkinetic}
 \ee
If the curvature scale $c_4$ (see (\ref{flatMsolution}) or (\ref{eq:IIAafsol})) of the throat is small compared to the volume of the compactification, a natural value of $u_{UV}$ would be the value $u_{UV} = c_4/\sqrt{\alpha'}$ at which the background (\ref{eq:IIAafsol}) opens up into a flat regime. 
 
Consider instead the case that the full values of $M_{pl}^2, f_{\phi}$ are of the order of $\delta M_{pl}^2, \delta f_{\phi}$ given above.  One consequence of (\ref{eq:throatkinetic}) is that $x = \lambda\theta/4\pi^2 N$ measures the canonically normalized axion field in Planck unit,
\[
x\equiv {\lambda\theta\over 4\pi^2 N}=\frac{f_a\theta}{\sqrt{2} M_{Pl}}\ .
\] 
As a consequence,  flattening of the axion potential happens exactly when a deviation of the canonically normalized axion field from the minimum is of order $M_{Pl}$.

\subsection{Axion-driven inflation}

The potential (\ref{thetapotential}) is an interesting candidate to drive inflation, as the flattening of the potential facilitates slow roll.  Furthermore, the monodromy may indeed allow an inflaton to travel over superPlanckian distances in the field space. The instabilities summarized in Table~\ref{default} happen at parametrically large values of $x$, and do not prevent transPlanckian excursions in the field space. This was at the root if the initial interest in axion monodromy inflation.

In our model, however, the flattening is more extreme than that discussed in \cite{Silverstein:2008sg,McAllister:2008hb}; the potential falls off as $A - B/\phi^6$ for canonically normalized inflaton $\phi$.   This prevents generation of the large B-mode polarization signal in this model---the axion potential saturates too fast and never reaches high enough values. The resulting inflationary  model is at the edge between large and small field models --- 50 e-foldings of inflation corresponds to a $\delta \phi \sim 2 M_{Pl}$ excursion in the field space. Such a model predicts a scalar index $n_s\sim 0.965$ (which is in perfect agreement with current WMAP data), and unobservably small tensor-to-scalar ratio, $r\sim 8\times 10^{-4}$.
 
On a more theoretical note, we can ask whether instabilities in Table~\ref{default} prevent the axion from reaching the regime of slow roll eternal inflation. The condition for a slow roll inflation to be eternal reads 
\be
\label{eternal_cond}
{V_x\over V^{3/2}}<{1\over\sqrt{2}\pi M_{Pl}^2}\;.
\ee
This condition gets satisfied at large $x$,
\be
{27\sqrt{3}\over 4}{N u_{UV}^2\beta^2\over \lambda^{3/2}}<x^7\;.
\ee
Thus, the eternally inflating regime is easier to reach at small values of  the UV cutoff $u_{UV}$. If we demand that the cutoff is still compatible with the existence of the throat at all values of $x$ down to $x=0$, 
it will be of order $u_{UV}={2\lambda\over 9\beta}K$, where $K$ is some number larger that one (say, $K=10$). In this case (\ref{eternal_cond}) reduces to
\be
{K^2N\sqrt{\lambda}\over \sqrt{3}}<x^7\;.
\ee
Interestingly, this is exactly the regime when the throat becomes unstable due to a  Casimir effect, (see the second line in Table~\ref{default}).\footnote{To avoid confusion, let us stress that this Casimir instability operates locally in the throat and does not get removed if $\beta$-modulus is stabilized.}  This does not completely rule out slow roll eternal inflation somewhere in the landscape, but we find the above coincidence intriguing.

\subsection{Axion strings}

Absent monodromy, $\theta$ can wind by $2\pi$ about an ``axion string'' defect.  If the axion has an instanton-generated potential, the string must climb over the $\Lambda^4 \cos(\phi/f)$ potential as it shifts by $2\pi$, and so it will form the boundary of a domain wall \cite{Vilenkin:1982ks}.  In the presence of monodromy, an interpolation of the axion by $2\pi$ will merely push the axion up a single branch. In a consistent configuration, one must jump to a lower branch during or at the end of the circuit in $\theta$.  Such a change of branches will, again, occur across a domain wall which ends on the axion string. We will make some general statements here and place the detailed calculations in Appendix B.

\begin{figure}[ht!]
\begin{center}
\includegraphics[scale=.4]{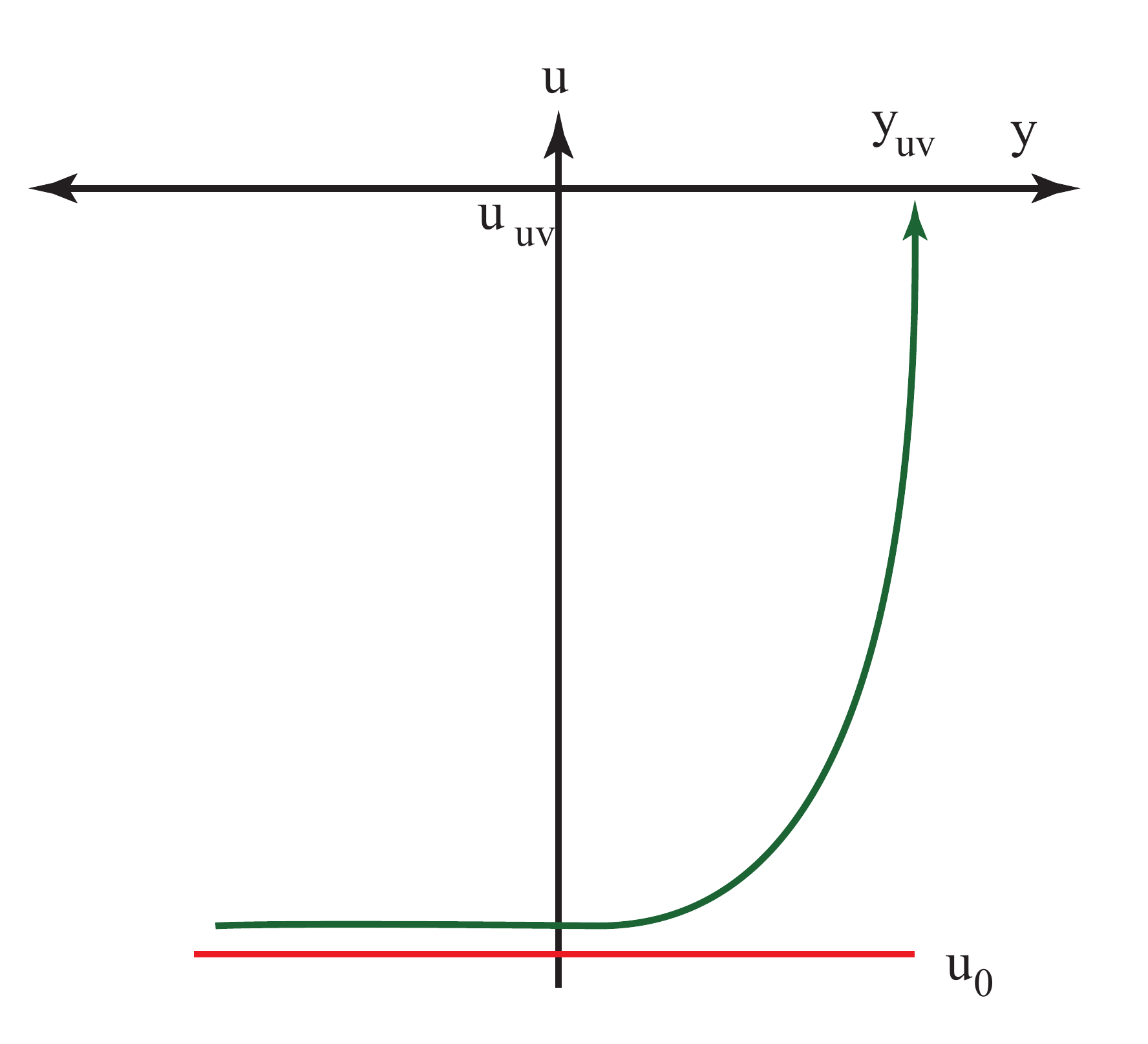}\includegraphics[scale=.4]{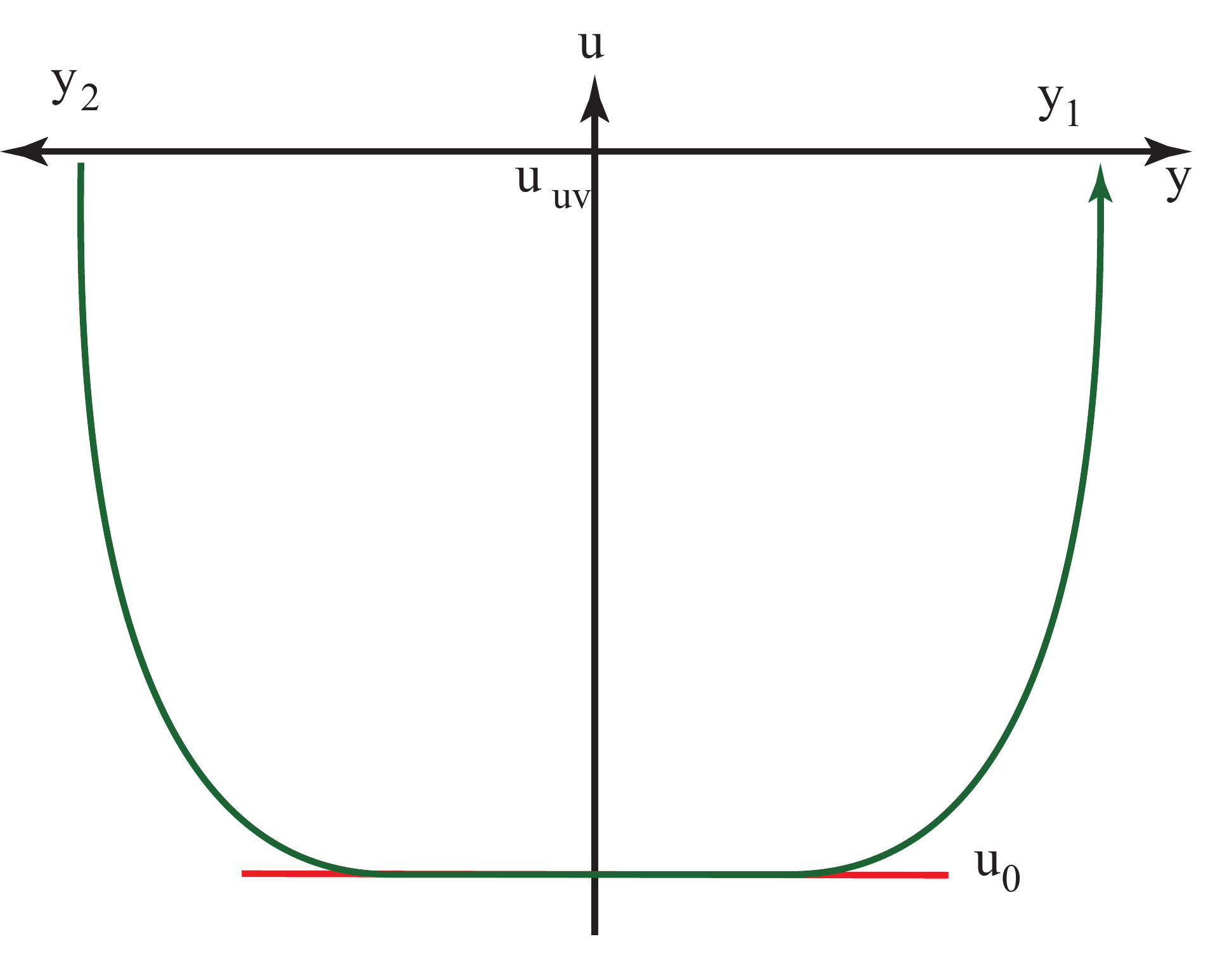}\end{center}
\vspace{-.5cm}
\caption{\label{stringdomain}On the left, the profile of the D6-brane in the $u-y$ plane, describing a string at $y_{UV}$ ending on a domain wall stretching to the left. On the right, the profile of a D6-brane in the $u-y$ plane, describing a domain wall bounded by an axion string at $y_1$ and an anti-string at $y_2$. }
\end{figure}

First, consider the ``deconfined phase'' (ignoring for now the instability at the singular horizon). The vacuum energy is $\theta$-independent and we do not expect the string to end on a domain wall. The candidate axion string is a D6-brane wrapping the $S^4$, running along $t, y_1, u$, at a point in $\xi, y_2, y_3$.   Because the D6-brane is magnetically charged with respect to the RR 1-form potential, $\int_{\chi} C^{(1)}$ will shift by $\sqrt{\alpha'}$ as we make a loop around the string in the $y_{2}, y_3$ plane.  Continuing this statement to large $u$ means that $\theta$ must shift by $2\pi$.  The pullback of $C^{(7)}$ onto this worldvolume vanishes, and after integrating the DBI action over the $S^4$, the DBI action is:
\be
	S_{deconfined} = \frac{N}{24\pi^3} \int dt dy_1 \int _0^{u_{UV}} du u\ ,
\ee
where $u_{UV}$ is a UV cutoff scale.  Thus, the contribution of the DBI action to the tension is
\be
	T_{deconfined} = \frac{N}{48\pi^2} u_{UV}^2
\ee
which is quadratically divergent, and $\Op(N)$.  The logarithmic IR divergence expected for global strings arises from the nonvanishing RR 2-form field strength sourced by the axion.

\begin{figure}[ht!]
\begin{center}
\includegraphics[height=3in]{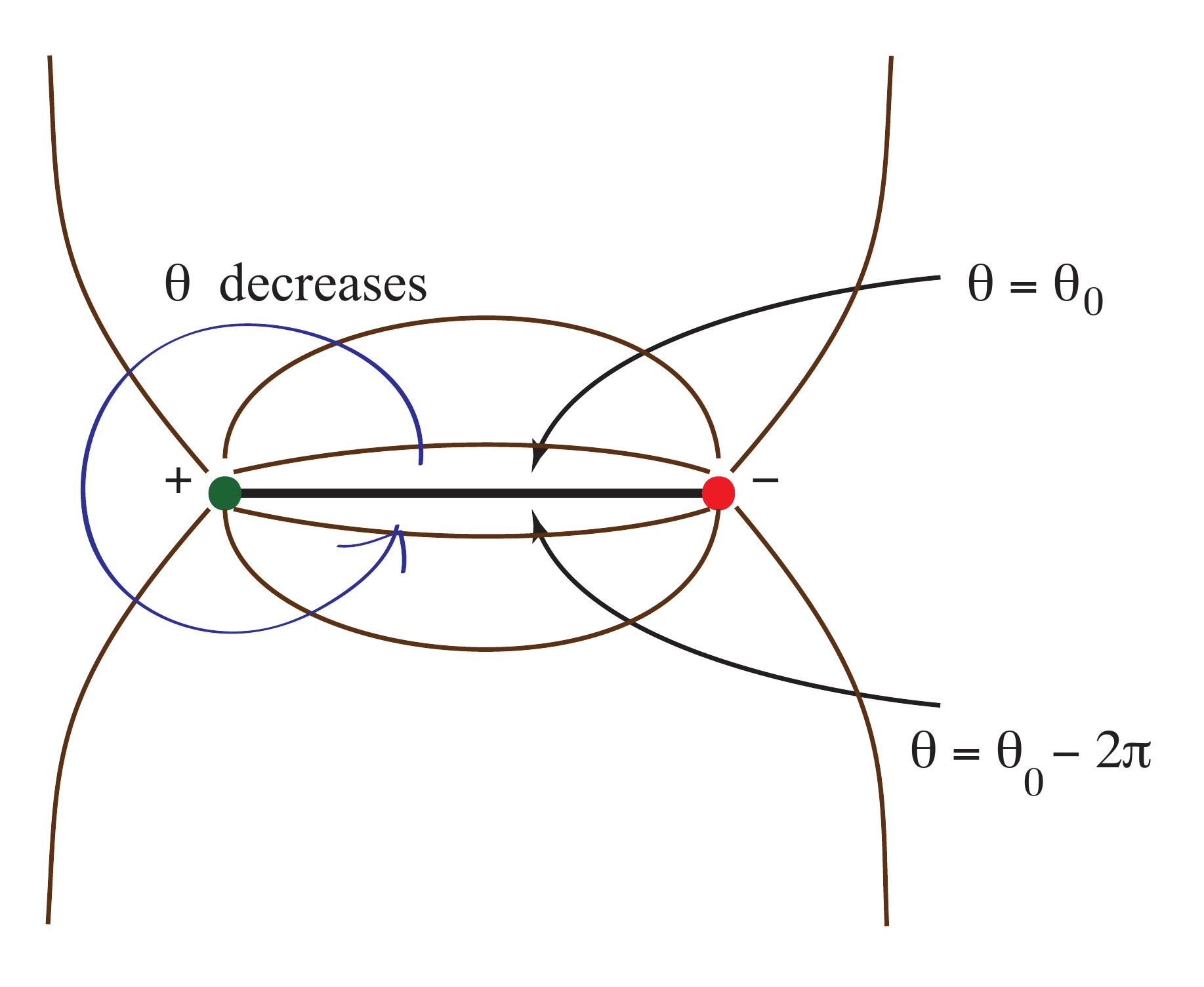}\end{center}
\vspace{-.5cm}
\caption{\label{dipole} The two points labeled "+" and "-" are the string and anti-string; the horizontal line joining them is the domain wall. The lines emanating from the string are contours of constant $\theta$; in this picture $\theta$ decreases as one moves counterclockwise about the string or clockwise about the anti-string.}
\end{figure}

At finite $u_0$, the string cannot end in the IR.  There are several possibilities:
\begin{itemize}
\item The string runs along $\chi = 0$, reaches the tip, and continues back to infinity along $\chi = \pi$.  This is an unstable string-anti-string pair.  
\item The D6-brane bends along the $u,y_2\equiv y$ direction, as shown in Fig. \ref{stringdomain}.  This describes a string ending on a domain wall.  If the axion begins at a long distance along a metastable branch, there is a force acting on the domain wall.  This can be seen either from the Chern-Simons part of the D-brane action, or by noting that the potential energy of the axion jumps across the wall. If the axion begins at the bottom of the potential, however, it will first evolve by $\delta\theta = \pi$ and then transition the next branch, evolving back to zero energy; this domain wall will feel no force.
\item There is a domain wall stretched between a string-anti string pair.  This is described by a D6-brane which follows a curve in the $u,y$ plane that asymptotes to large $u$ at two values of $y$ separated by $\Delta y$, as seen in Fig. \ref{stringdomain}.  As $\Delta y$ increases, the minimum value of $u$, describe by a domain wall, approaches $u = u_0$.  For larger values of $\Delta y$, the D6-brane will hang down to $u = u_0$ run along there as a domain wall, and then  run back up to large $u$.  If the axion is high up along a metastable branch, this configuration will again feel a force perpendicular to the domain wall. One can deduce the topology of the lines of constant axion, shown in Fig (\ref{dipole}).
\end{itemize}

A more detailed analysis of the DBI-CS action of the D6-brane is given in Appendix B.

\section{Conclusions}

We take two main lessons from this work.

\begin{itemize}
\item At large values of $\theta$, the confinement scale $\Lambda(\theta)$ and glueball masses of our strongly coupled field theory decrease.  Since this is the only scale in the problem, we expect $\p_{\theta} E(\theta) \sim \Lambda(\theta)^4$, and the potential flattens. This is the 4d field theory analog of the flattening of the axion monodromy potential seen in string theory compactifications \cite{Dong:2010in}.
\item The metastable branches at large $\theta$ eventually become unstable, as found in a different context in \cite{Dubovsky:2010je} and as conjectured for softly broken ${\cal N} = 1$ gauge theories in \cite{Shifman:1998if}.  The fact that the potential barrier between branches shrinks at large $\theta$ is consistent with the lowering of the dynamical scale $\Lambda(\theta)$ of the theory. 
\end{itemize}

There are a number of next steps worth pursuing, both formal and phenomenological.

In formal terms, it would be interesting to explore axion monodromy in other field theories. For example, we could consider beginning with a Euclidean black M2-brane, followed by a Euclidean rotation and analytic continuation to get a holographic dual of a $1+1-d$ theory analogous to ours.
It would be even more interesting to find calculable examples in which the UV and IR dynamics are of the same dimension, such as in softly broken sypersymmetric gauge theory \cite{Shifman:1998if}. In this regard, two-dimensional models may be interesting.  The Schwinger model -- two-dimensional electrodynamics coupled to charged fermions -- exhibits this behavior; the axion couples to the topological term $\int d^2x F_{01}$. The energy increases quadratically on any given branch, and the branch can change via the pair production of charged electrons.  One does not see the flattening of the potential observed here since the theory is too simple -- the gauge field has no dynamics.  A richer possibility would be the $\mathbb{CP}^n$ model in two dimensions, which has a topological term and instantons, and at low energies reduces to abelian gauge theory coupled to charged matter \cite{DAdda:1978uc,Witten:1978bc}.  
 
For cosmological purposes, we would like to better understand the embedding of this theory into a full string compactification -- most importantly, we would wish to see if our inflation scenario survives moduli stabilization.
Another important avenue is to apply the insights we gain here to other realizations of mondromy inflation -- for example, to study the possibility of the decay of the metastable branches in the explicit string theory models  described in \cite{Silverstein:2008sg,McAllister:2008hb}.

\vskip .5cm

\centerline{\bf Acknowledgements}
\vskip .5 cm
We would like to thank Raphael Flauger, Shamit Kachru, Ami Katz, Matthew Kleban, and Eva Silverstein for helpful discussions.  The work of A.L. is supported by DOE Grant DE-FG02-92ER40706. M.M.R. is supported by the Simons Postdoctoral Fellowship Program.

\appendix
\section{Identifying scales across frames}

In the draft we made the perhaps puzzling statement that the 11d Planck scale $\ell_{11}$ measured in the Einstein frame of 11d supergravity was equal to the string scale $\sqrt{\alpha'}$ measured in the string frame of type IIA string theory in ten dimensions. The reader familiar with the classic formula $\ell_{11} = g_{s}^{1/3} \sqrt{\alpha'}$ may be confused about this (as were we at first).  Nonetheless, both statements are in fact true, as we will now explain.

Begin with the Einstein action in 11d Einstein frame:
\be
	S_{11d} = \frac{1}{(2\pi)^8 \ell_{11}^9} \int d^{11} x \sqrt{g_{11}} R[g_{11}]
\ee\label{eq:mtheoract}
Now compactify to ten dimensions by setting $x_{10}$ to be  a circle with proper radius $R_{11} = e^{2\phi_0/3} \ell_{11}$, using the metric
\be
	ds_{11}^2 = e^{4\phi_0} d(x^{11})^2 + e^{-2\phi_0/3} g_{10,\mu\nu} dx_{10}^{\mu}dx_{10}^{\nu}
\ee
In the reduction of M theory to type IIA string theory, $g_{10}$ is the string frame in ten dimensions, and $e^{\phi_0} = \equiv g_s$ is the string coupling constant. The 10d action under this reduction is
\begin{eqnarray}
	S & = & \frac{2\pi R_{11}}{(2\pi)^8\ell_{11}^9} \int d^{10} x \sqrt{g_{10}e^{20\phi_0/2}}
		e^{2\pi_{0}/3} R[g_{10}] \nonumber\\
	& = & \frac{1}{(2\pi)^7\ell_{11}^8} \int d^{10}x \sqrt{g_{10}} e^{-2\phi_0} R[g_{10}]
\end{eqnarray}
This is the type IIA action in string frame if we set $\ell_{11} = \sqrt{\alpha'}$. We get a similar story if we compare the Nambu-Goto action for the M-theory membrane to the action for the D2-brane in type IIA string theory.

The point is that in the previous paragraph, $\ell_{11}$ is the 11d Planck scale measured in the Einstein frame.  The same dimensionful number, arising as the result of a measurement in string frame, is the string scale $\sqrt{\alpha'}$.   When we write a relation such as $\ell_{11} = g_{s}^{1/3} \sqrt{\alpha'}$ we are comparing two dynamical scales {\it measured in the same frame}. For example, consider the energy of an oscillator mode of an 11d membrane transverse to $x_{11}$.  Following the procedure above, the action is
\begin{eqnarray}
	S &=& \frac{1}{(2\pi)^2\ell_{11}^3} \int d^3\sigma \sqrt{- \det g_{11,\mu\nu} \p_{\alpha} X^{\mu} \p_{\beta} X^{\nu}}\nonumber\\
	& = & \frac{e^{-\phi_0}}{(2\pi)^2 \ell_{11}^3} \int d^3\sigma \sqrt{- \det g_{10,\mu\nu} \p_{\alpha} X^{\mu} \p_{\beta} X^{\nu}}
\end{eqnarray}
where $\sigma^{\alpha = 0,1,2}$ are the worldsheet coordinates and $X^{mu}$ the target space coordinates. Again, this is the D2-brane action, (moduli the gauge field which is the dual of motion along $x_{11}$) as predicted by M-theory-type IIA duality, with the appropriate tension $T_{D2} = 1/)g_s (\alpha')^{3/2}$, if we set $\ell_{11} = \sqrt{\alpha'}$.  If we adopt an operational definition of ``the 11-d Planck scale'' as $L_{11} \equiv T^{-1/3}$ -- a scale we can measure -- then in string frame we will find that $L_{11} = g_s^{1/3} \sqrt{\alpha'}$.  If we measure this same scale using the 11d metric we will find $L_{11} = \ell_{11}$.  

As a check on this, M-theory-type IIA duality identifies the M2-brane wrapped on $x_{11}$ as the fundamental type IIA string.  We can do this by setting $\sigma^2 = x^{11}$, so that
\begin{eqnarray}
	S & = & \frac{1}{(2\pi)^2\ell_{11}^3} 2\pi R_{11} \int d^2 x e^{-2\phi_0/3} \sqrt{- \det g_{10,\mu\nu} \p_{\alpha} X^{\mu} \p_{\beta} X^{\nu}}\nonumber\\
	& = & \frac{1}{2\pi\ell_{11}^2} \int d^2 x \sqrt{- \det g_{10,\mu\nu} \p_{\alpha} X^{\mu} \p_{\beta} X^{\nu}}
\end{eqnarray}
which is the Nambu-Goto action for the fundamental string when $\ell_{11} = \sqrt{\alpha'}$.  The ``string scale'' in the 10d string frame can be defined operationally as the energy scale of a string oscillator mode; the ratio of this scale and the 11d scale, when both are defined operationally by a string scale observer, we find $L_{11}/\sqrt{\alpha'} = g_s^{1/3}$ as expected.  If we measured the oscillations of this object in the M theory metric, the tension is $T_{s} \sim R_{11}/(2\pi \ell_{11}^3)$; the ratio between the energy $T_{s}^{1/2}$ of oscillator modes of this object, as measured in 11d Einstein frame, and the energy $1/\ell_{11})$ of oscillator modes of the M2-brane as measured in 11d Einstein frame, we find
\be
	\frac{1}{\ell_{11} T_s^{1/2}} \sim \sqrt{\frac{\ell_{11}}{R_{11}}} \sim e^{\phi_0/3} = g_s^{1/3}
\ee
Thus the dimensionless ratio of two scales measured in a fixed frame is independent of frame, as expected.

\section{Detailed analysis of axion string-domain wall configurations}

In this appendix we give a more detailed explanation of the results cited in \S5.3, by studying the action for the D6-brane in the probe limit.  This limit is valid if $x$ is large, so that the jump in the RR 2-form flux across the domain wall is small compared to the total 2-form flux.

Consider a D6-brane wrapping the $S^4$ of our geometry, filling out $t,y_1$, and sitting at $y_2 = 0$.  We will study the action for profiles $u(y\equiv y_2$. The action is the sum of the Dirac-Born-Infeld (DBI) action, and of the Chern-Simons (CS) action describing the coupling of the D6-brane to the 7-form RR potential that is dual to the RR 1-form. This seven-form RR potential can be written as:
\be
C^{(7)}=- \ap^{7/2}\frac{16\pi \lambda^2Nx}{3^5\beta^4(1+x^2)^4}y_3~dt\wedge dy_1\wedge dy_2\wedge \epsilon^{(4)}_\Omega
\ee
The full DBI+CS action in this case is:
\begin{eqnarray}
	S & = & \frac{N}{24\pi^2} \left(\frac{2}{\lambda\beta}\right)^{1/2} \int dt dy_1 dy
		u^{5/2} H_0(u) \sqrt{1 + \frac{\lambda\beta}{2 u^3 f} u_y^2} \nonumber\\
	& & \ \ \ \ + \frac{\lambda^2 N x}{4 3^5 \beta^4 (1 + x^2)^4} \int dt dy_1 dy_2\times y_3(t,y_1,y)
\end{eqnarray}
The Chern-Simons form once again exerts a force transverse to the domain wall, as stated in \S5.3.

Since $S_{DBI} = \int d^3y L$ is $y$-independent, the following "conserved" quantity is $y$-independent:
\begin{eqnarray}
	E_y & = & u_y \frac{\delta S}{\delta u_y} - L\nonumber\\
	& = & - \frac{N}{24\pi^2} \left(\frac{2}{\lambda\beta}\right)^{1/2} \frac{u^{5/2} H_0}{\sqrt{1 + \frac{\lambda\beta}{2u^3 f} (u_y)^2}}
\end{eqnarray}
Solving for $u_y$, we find:
\be
	u_y = \sqrt{\left(\frac{2 u^3 f}{\lambda\beta}\right) \left(\frac{N^2 u^5 H_0^2}{288 \pi^2 \lambda\beta E_y^2} - 1 \right)} \label{eq:sdprofile}
\ee

At large $u$, the solution scales as $u_y \sim A u^4$, so that $y \sim B/u^3$ and runs off to $u = \infty$ as $y \to 0$ (where we have fixed the integration constant of the differential equation for $u$. )  In this regime, the solution looks like the axion string.

If $E_y$ is sufficiently large that the second term in parenthesis under the square root of (\ref{eq:sdprofile}) has a zero for $u_E \gg u_0$, it will at most be a simple zero at some value $y = y_E$, so that $u \sim U_E + b (y - y_E)^2$.  At $y = y_0$, the solution can be joined onto that of a domain wall emanating from an antistring, described by a D6-brane which reaches the boundary $u = \infty$ at $y = 2 y_E$, as in Fig. (\ref{stringdomain})  In this regime, scaling shows that $y_E \sim 1/\sqrt{E_y}$, so that the strings get closer and closer together and the D6-brane turns in the $u$ direction at large and large $u$, as $E_y \to \infty$. One side of the domain wall will see larger values of $\theta$, so that we expect a force to act on the domain wall, consistent with the force seen in the Chern-Simons coupling.

As we lower $E_y$, the largest zero in the second term approaches the zero in $f$.  At this degenerate point, $u_y \sim b(u - u_0)$, so that $u - u_0 \sim e^{- by}$.  In this limit, the string-anti string pair moves off to infinite distance. Call this value of $E_y = E_{y,0}$.  

For smaller $E_y < E_{y,0}$, the largest zero inside the square root remains a single zero at $u = u_0$.  The D6-brane hits $u_0$ at a finite value of $y_0$, behaving as $u - u_0 \sim c (y - y_0)^2$. As $E_y$ gets smaller, the curve in the $u -y$ plane becomes more vertical (more and more parallel to the $u$ axis).  At this point, the configuration can be smoothly joined onto a second D6-brane with the same value  of $E_y$ the reaches back out to infinity; the separation then becomes smaller as a function of $E_y$.  For fixed separation between the string and anti-string at infinity, this configuration will have large energy than the configuration with $E > E_{y,0}$.  This story, with two solutions at a given string separation, is somewhat reminiscent of the holographic dual of quark-anti quark pairs in finite-temperature ${\cal N}=4$ gauge theory \cite{Rey:1998bq,Brandhuber:1998bs,Brandhuber:1998er}.

\eject
\bibliographystyle{utphys}
\bibliography{dlrrefs}

\end{document}